\documentclass[acmsmall]{acmart}

\usepackage{algorithmic}
\usepackage{xcolor}
\usepackage[linesnumbered,ruled,vlined]{algorithm2e}

\SetCommentSty{mycommfont}
\SetKwInput{KwInput}{Input}                
\SetKwInput{KwOutput}{Output}              

\newcommand{\expnumber}[2]{{#1}\mathrm{e}{#2}}

\usepackage{multirow}
\usepackage{array}
\usepackage{subcaption}

\usepackage[normalem]{ulem}
\useunder{\uline}{\ul}{}

\usepackage{framed,enumitem} 

\newcommand{\ie}{\textit{i.e.}, } 

\newcommand{\naive}{na\"\i{}ve } 

\newcommand{\workname}{LERG}

\AtBeginDocument{%
  }

\setcopyright{acmlicensed}
\copyrightyear{2018}
\acmYear{2018}
\acmDOI{XXXXXXX.XXXXXXX}

\acmConference[Conference acronym 'XX]{Make sure to enter the correct
  conference title from your rights confirmation emai}{June 03--05,
  2018}{Woodstock, NY}
\acmISBN{978-1-4503-XXXX-X/18/06}

\begin{document}

\title{Lightweight Embeddings with Graph Rewiring for Collaborative Filtering}

\author{Xurong Liang}
\email{xurong.liang@uq.edu.au}
\orcid{0000-0002-3458-3887}
\affiliation{%
  \institution{The University of Queensland}
  \city{Brisbane}
  \country{Australia}
}

\author{Tong Chen}
\email{tong.chen@uq.edu.au}
\orcid{0000-0001-7269-146X}
\affiliation{%
  \institution{The University of Queensland}
  \city{Brisbane}
  \country{Australia}
}

\author{Wei Yuan}
\email{w.yuan@uq.edu.au}
\orcid{0000-0002-9400-842X}
\affiliation{%
  \institution{The University of Queensland}
  \city{Brisbane}
  \country{Australia}
}

\author{Hongzhi Yin}
\email{h.yin1@uq.edu.au}
\orcid{0000-0001-7269-146X}
\authornote{Hongzhi Yin is the corresponding author.}
\affiliation{%
  \institution{The University of Queensland}
  \city{Brisbane}
  \country{Australia}
}

\renewcommand{\shortauthors}{Xurong Liang et al.}

\begin{abstract}
GNN-based recommender systems have become increasingly popular in academia and industry due to their ability to capture high-order information from user-item interaction graphs. However, as recommendation services scale rapidly and their deployment now commonly involves resource-constrained edge devices, GNN-based models face significant challenges, including high embedding storage costs and runtime latency from graph propagations.
Our previous work, LEGCF, effectively reduced embedding storage costs but struggled to maintain recommendation performance under stricter storage limits. Additionally, LEGCF did not address the extensive runtime computation costs associated with graph propagation, which involves heavy multiplication and accumulation operations (MACs). These challenges consequently hinder effective training and inference on resource-constrained edge devices.
To address these limitations, we propose \textbf{L}ightweight \textbf{E}mbeddings with \textbf{R}ewired \textbf{G}raph for Graph Collaborative Filtering (\workname), an improved extension of LEGCF. \workname \text{} retains LEGCF’s compositional codebook structure but introduces quantization techniques to reduce the storage cost of embedding weights, enabling the inclusion of more meta-embeddings within the same storage constraints for improved model expressiveness. To optimize graph propagation for edge devices, we pretrain the quantized compositional embedding table using the full interaction graph on resource-rich servers, after which a fine-tuning stage is engaged to identify and prune low-contribution entities via a gradient-free binary integer programming approach, constructing a rewired graph that excludes these entities (\ie user/item nodes) from propagating signals.
The quantized compositional embedding table with selective embedding participation and sparse rewired graph are transferred to edge devices which significantly reduce computation memory and inference time.
Experiments on three public benchmark datasets, including an industry-scale dataset, demonstrate that \workname \text{} achieves superior recommendation performance while dramatically reducing storage and computation costs for graph-based recommendation services.

\end{abstract}



\received{20 February 2007}
\received[revised]{12 March 2009}
\received[accepted]{5 June 2009}

\maketitle

\section{Introduction} \label{sec:intro}
Recommender systems have been widely deployed in our daily lives to help people make better decisions when shopping, listening to music, watching movies, and more \cite{linden2003amazon, covington2016deep, fan2019graph,wang2018billion, zheng2025can}. Early recommender systems directly compute the dot product between representation vectors (i.e., embeddings) of users and items for recommendation score calculation \cite{rendle_factorization_2010}. Following the recent surge of study in graph neural networks (GNNs) \cite{liu2022survey, wu2022graph},  GNN-based recommender systems, which leverage the power of GNNs to learn user and item representations, have been a popular deployment choice and achieved remarkable success in both academia \cite{he_lightgcn_2020, wang_neural_2019, yu2022graph, yu2023xsimgcl} and industry \cite{gurukar2022multibisage,pal2020pinnersage} due to their strong capability in capturing and propagating user/item nodes' 
collaborative signals across their neighbors for interaction modeling \cite{wu2022graph}.

As the need for scalability and decentralized services arises, researchers have shifted their attention to optimizing the training efficiency and model deployment~\cite{chen_learning_2021, xia2023towards}, especially on edge devices. Traditionally,
recommender systems follow a cloud-based deployment mechanism, where the recommendation model is wholly trained and stored on a resource-rich server. The edge devices only act as agents that query and receive the recommendation results from the central server via a communication link \cite{yin2024device}. This working mechanism has been criticized for being overly reliant on internet access and central servers as well as the high response latency caused by communication overhead \cite{yin2024device}. To this end, recent studies \cite{tran2025thorough, liu_learnable_2021, chen_learning_2021, xia2023towards,qu2023continuous,qu2024budgeted, mairittha2020improving, lyu_optembed_2022, liang2023learning, long2023decentralized,zheng2024personalized} have been focusing on deploying lightweight recommender systems on resource-constrained edge devices directly. 
A general deployment pipeline for this kind of work involves initially training a full-scale recommender system on a central server, followed by compressing and deploying a lightweight version to edge devices to accommodate their hardware constraints.
The on-device model is expected to perform inference fully on board instead of querying the central server in real-time \cite{yin2024device}. 
Considering the versatility and robust performance of GNN-based recommender systems in cloud-based settings, how to efficiently train and deploy GNN-based recommender systems on edge devices raises extensive demands~\cite{wu2022graph}. However, the transition to edge-based recommendation services faces two major challenges.

The first major challenge is the substantial storage cost of the recommendation model caused by the heavily parameterized embedding table, a problem widely acknowledged in embedding optimization research~\cite{li2024embedding}. While many approaches attempt to address this, they often fail to achieve a satisfactory balance between storage cost and recommendation performance. For instance, dimension search algorithms \cite{liu_automated_2020, zhao_autoemb_2020, qu2023continuous, qu2024budgeted,chen_learning_2021, lyu_optembed_2022} determine the optimal embedding dimension size for each user/item. However, the objectives of these methods typically prioritize maintaining model accuracy and have weak or none model size constraints, making it difficult to adhere to tight storage budgets. Pruning-based methods \cite{liu_learnable_2021, qu2022single, yao2023razor} are effective in reducing storage by eliminating redundant or irrelevant embedding parameters, but this is at the cost of significantly reducing the number of usable embedding dimensions available to each user/item. Under strict storage constraints, this aggressively weakens the embeddings' expressiveness and can significantly hinder recommendation performance. Moreover, both dimension search and pruning-based approaches involve resource-intensive and time-consuming search or retraining processes, limiting their feasibility for large-scale deployment.
The other promising optimization strategy, compositional embedding \cite{weinberger_feature_2009, shi_compositional_2020, liang2023learning, liang2024lightweight}, maps multiple users/items to shared meta-embeddings to reduce the total number of embeddings stored. Our prior work, \textbf{LEGCF} \cite{liang2024lightweight}, exemplifies this approach. LEGCF introduces a compact, dense codebook combined with a scalable, sparse assignment weight matrix, generating entity\footnote{In this paper, we term both users and items as entities for convenience.} embeddings by a unique combination of two meta-embeddings with assignment weights. Additionally, the assignment weight matrix is learnable, allowing dynamic updates to better capture evolving entity representations.
Despite significantly reducing storage costs, LEGCF still faces challenges in balancing storage constraints with recommendation performance. 
Specifically, under strict storage constraints, LEGCF may encounter severe "collisions" of meta-embeddings when using a codebook to represent entity embeddings, i.e., a large number of entity embeddings may be represented by the same meta-embedding, resulting in the degradation of recommendation performance.

The second major challenge in deploying GNN-based recommender systems on resource-constrained edge devices is the high runtime computation cost.  GNN-based recommender systems rely extensively on repetitive graph propagation operations to exchange collaborative signals between entities and encode them into their embeddings \cite{wu2022graph}. For simplicity, consider a graph convolution network (GCN) \cite{kipf2017semi, he_lightgcn_2020} without nonlinear activation, where graph propagation is effectively a matrix multiplication between the graph adjacency and entity embedding matrices. The computation cost of matrix multiplication is typically measured by the number of multiplication and accumulation operations (MACs) it requires. 
As an example, our largest experimental dataset iFashion (see Section~\ref{sec_dataset}) entails over 2.6 million interactions among a total of 2 million users and items.  Propagating embeddings across such a graph using a GCN requires 10 billion MACs in a single pass, even before taking multiple convolution layers into account. 
Although some embedding compression techniques can compact GNN-based recommender systems to fit on a device~\cite{li2024embedding}, they often overlook the runtime computation costs. As a result, the accumulative MACs of GNN-based recommendation models not only hurt inference efficiency, but also negate the potential of performing on-device fine-tuning to handle new interactions over time. Our prior work, LEGCF, for example, constructs an expanded interaction graph and propagates both entity and codebook embeddings during the assignment weight update phase. This process imposes a substantial runtime computation cost, hindering deployment on edge devices. 
Several general-purpose techniques \cite{wu2019simplifying, zeng2020graphsaint, hamilton2017inductive, chen2020scalable, qu2023tt, gupta2024graphscale} address this issue by performing mini-batch training during graph propagation, using sampled neighbors \cite{hamilton2017inductive, gupta2024graphscale} or subgraphs \cite{zeng2020graphsaint, chiang2019cluster, qu2023tt} for feature propagation in each batch. However, the sampling process may take a considerably long time for large-scale graphs. Also, the sampling heuristics play an important role in determining entity embedding quality, and the accuracy of GNN-based recommender systems may be heavily affected when the graph-propagated entity embeddings do not correctly reflect the neighbor relations and collaborative semantics.

In this paper, we extend our previous work~\cite{liang2024lightweight} and propose an efficient framework designed for lightweight GNN-based recommendation, namely \textbf{L}ightweight \textbf{E}mbeddings with \textbf{R}ewired \textbf{G}raph for Graph Collaborative Filtering (\textbf{\workname}). 
To address the substantial storage cost of the embedding table, we build upon the design of LEGCF to employ a compositional embedding table comprising a dense codebook and a highly sparse assignment weight matrix. 
We identified that LEGCF’s performance degradation stems from its reliance on conventional 32-bit floating-point precision for meta-embeddings, which consumes significant storage. Under stringent storage constraints, reducing the number of meta-embeddings in the codebook becomes necessary, leading to diminished compositional diversity and compromised entity embedding uniqueness, ultimately impacting model accuracy. In this regard, quantization \cite{wang2023bitnet, nagel2021white, guan2019post, ma2024era, sun2020ultra} offers an effective solution to reduce the storage size of individual meta-embedding elements. Widely validated in the field of large language models (LLMs) \cite{wan2023efficient, chen2024efficientqat, ma2024era, xi2023training, wang2023bitnet}, quantization preserves model weight quality while accelerating training and inference. 
While one can apply quantization techniques to the full embedding table directly \cite{guan2019post, xu2021agile, li2023adaptive} to optimize the entity embeddings, an aggressively low precision is often used to meet a tight storage constraint, which heavily impedes the fidelity of entity embeddings.
Therefore, instead of applying quantization techniques to the full embedding table, \workname \ quantizes each full-precision meta-embedding in the dense codebook into a low-bit integer representation. This allows the embedding table to accommodate more meta-embeddings within the same storage space, reducing the risk of hash collisions and enhancing the uniqueness of entity embeddings.

To enable computationally efficient graph propagation on resource-constrained edge devices, we first pretrain the compositional embedding table on resource-rich servers using the full user-item interaction graph. Next, we evaluate each entity’s contribution to collaborative signal propagation by framing this as a gradient-free binary integer programming problem. Entities with the lowest contributions are identified, and a rewired graph is generated that excludes these entities from propagating collaborative signals to their neighbors.
The pretrained, quantized compositional embedding table is then efficiently fine-tuned using the rewired graph.
To further reduce peak memory consumption and interference during fine-tuning, we disable the embeddings of pruned entities from refining the quantized compositional embedding table. Instead, these embeddings are drawn from a small set of placeholder meta-embeddings derived from the pretrained graph-propagated entity embeddings. Consequently, only a subset of entity embeddings participates in graph propagation, further lowering runtime computation costs. For deployment, only the pretrained quantized compositional embedding table, the rewired graph and the set of placeholder meta-embeddings along with its assignment vector for pruned entities are transmitted to resource-constrained edge devices. This approach minimizes the storage cost of the embedding table while significantly reducing the number of MACs required for graph propagation, alleviating computational complexity.

We summarize our main contributions as follows:
\begin{itemize}
    \item We analyze the features and drawbacks of common embedding optimization methods and graph computation improvement work. We reinforce the importance of controlling the storage cost of the embedding table and the runtime memory consumption of GNN-based recommender systems on resource-constrained edge devices.
    \item We put forward a substantially improved GNN-based lightweight recommender system, namely \workname, as an extension to our prior work LEGCF. 
    Essentially, LEGCF proposes using a dense codebook for entity embedding storage, in which all meta-embeddings are stored in 32-bit floating-point precision. A highly sparse meta-embedding assignment matrix is learned, controlling the meta-embedding composition of each entity. In LERG, we further reduce the storage cost of meta-embeddings in the codebook by applying quantization techniques. Under the same storage budget, the quantized compositional codebook allows more meta-embeddings to be stored, thus massively improving entity composed embedding uniqueness. Meanwhile, to address the runtime memory overhead of LEGCF, LERG uses a fixed assignment matrix, and a graph rewiring method is proposed to allow efficient propagation on a pruned interaction graph. The design of the rewired sparse graph significantly reduces runtime memory usage and computational time for graph propagation, enabling efficient fine-tuning and inference on edge devices.
    \item We conduct extensive experiments on three public benchmark datasets, including one on an industry scale. The experimental results verify the state-of-the-art recommendation performance of LERG compared with recent lightweight baselines.
\end{itemize}

\section{Related Work} \label{sec:related_work}
In this section, we outline work that is relevant to our research. We separate the related work into three categories: embedding storage optimization, server-edge computing paradigm and graph computation complexity optimization. 

\textbf{Embedding Storage Optimization.} To reduce the storage cost of the embedding table, most dimension search algorithms \cite{zhao_autoemb_2020, liu_automated_2020, qu2023continuous, qu2024budgeted, qu2024scalable,joglekar_neural_2020} formulate automated machine learning (AutoML) processes to select optimal dimension sizes from a predefined candidate size set.  Some others \cite{chen_learning_2021, lyu_optembed_2022} employ evolutionary search to identify an optimal embedding structure. Pruning-based methods \cite{liu_learnable_2021, liang2023learning, qu2022single, yao2023razor, li2023dual,qu2024sparser,tran2025device} instead define a pruning constraint on the embedding table. As the value update process continues, elements considered redundant or unimportant will be nullified to save space. The compositional embedding scheme is another major direction for embedding storage optimization. These methods reduce the number of embeddings to be stored on devices by letting multiple entities share the same set of embeddings (\ie meta-embeddings). The most common meta-embedding assignment strategy is to design a set of fixed hash functions \cite{weinberger_feature_2009, shi_compositional_2020, li_lightweight_2021, zhang_model_2020}. LEGCF \cite{liang2024lightweight} introduces a learnable meta-embedding assignment scheme to improve flexibility in generating entity embeddings. Our work roots from the backbone of LEGCF but avoids the assignment update process as it requires the involvement of meta-embeddings in graph propagation, which introduces additional graph computation overhead. Locality-sensitive hashing (LSH) can also be deployed for meta-embedding assignment \cite{desai2021semantically, das2007google}. Tensor train (TT) decomposition \cite{yin2021tt, qu2023tt, yin_nimble_2022, xia2022device, wang_next_2020} is another form of compositional embedding technique which forms entity embeddings as a product of tensors. Some researchers propose to learn several codebooks \cite{lian_lightrec_2020, zhang2022anisotropic, liu2020online, liu2021online, lian2020product, xia2023efficient}, each entity embedding is composed of one codeword drawn from each codebook. This strategy is termed vector quantization \cite{yin2024device} and can be regarded as a form of compositional embedding technique as well.

\textbf{Server-Edge Computing Paradigm.}
The main usage of server-edge computing paradigm is in federated learning \cite{liu2023federated, liu2023privaterec, zhang2023lightfr}, in which the edge devices first train a local model using a small local dataset, the parameters of locally trained models are then uploaded to central servers to perform a global aggregation \cite{yin2024device}. After that, the aggregated parameters are distributed to edge devices for inference. Such a training paradigm avoids the exposure of local data to the public, thus providing data security and user privacy. Another branch of research lines in on-edge fine-tuning, wherein a foundation model is first pretrained on resource-rich servers. After that, the foundation model is fine-tuned on edge devices to fit downstream tasks \cite{wang2024end}. \cite{wang2022fast, yan2022device} conducts fine-tuning on edge devices to update the entire pretrained foundation model, which elevates concerns on feasibility due to limited hardware specifications. \cite{mairittha2020improving} proposes to only update a part of the pretrained model to reduce the computation overhead on edge devices. Patch-learning \cite{yao2021device, long2024diffusion} is another direction that introduces a small patch model on top of the pretrained model. In the fine-tuning stage, only the weights of the patch model are updated to adopt the model to local deployment. Our work fine-tunes the quantized compositional embedding table by leveraging only a part of entity embeddings to control the peak runtime memory usage. The propagation graph is also rewired to reduce the graph computational complexity.

\textbf{Graph Computation Complexity Optimization.} To reduce the graph computation complexity, a node sampling strategy is commonly used, which samples a subset of nodes from the neighborhood of the target node to perform graph propagation at each mini-batch \cite{ding2022data}. Initially, the uniform sampling strategy is implemented \cite{hamilton2017inductive, chen2018stochastic}. To further optimize the sample time for large graphs, the sampling is later conducted in a layer-wise manner \cite{chen2018fastgcn, huang2018adaptive, zou2019layer}. The sampling strategy can also be conducted at the graph level so that a large graph is split into multiple subgraphs, each of which contains sampled nodes and edges. GraphSAINT \cite{zeng2020graphsaint} is a typical example of using the graph sampling technique to bring down the computation cost for graph propagation. Other methods \cite{chiang2019cluster, lin2020pagraph} seek to partition the original large graph into multiple subgraphs so that graph propagation is only conducted within each partition, or leverage the fast Fourier transform algorithm to map the graph vertex space into Fourier domain for efficient cross-correlation computation \cite{zhang2021multi}. Another research direction for graph computation complexity optimization focuses on graph sparsification, in which the number of connected edges in the graph is lowered so that the graph computation overhead, runtime memory consumption and communication latency can be improved when performing graph propagation over the entire graph \cite{liu2022survey}. There are methods, such as DropEdge \cite{rong2019dropedge}, which adopt handcrafted heuristics to decide the edges to be dropped. However, the heuristics for edge removal can be hard to tune and yet the model accuracy will be highly reliant upon the quality of the sparsified graph. Some other techniques \cite{yang2019topology, chen2021unified, zheng2020robust} construct an optimization constraint to learn a sparsified version of the interaction graph, despite this strategy often introducing additional trainable parameters. In our work, we leverage the pretrained graph-propagated embeddings to identify entity contributions in graph propagation and remove edges of entities with limited contributions, which can be regarded as a graph sparsification technique as well.

\section{Method} \label{sec:method}
\begin{figure}[t!] 
    \centering
            \includegraphics[width=\textwidth]{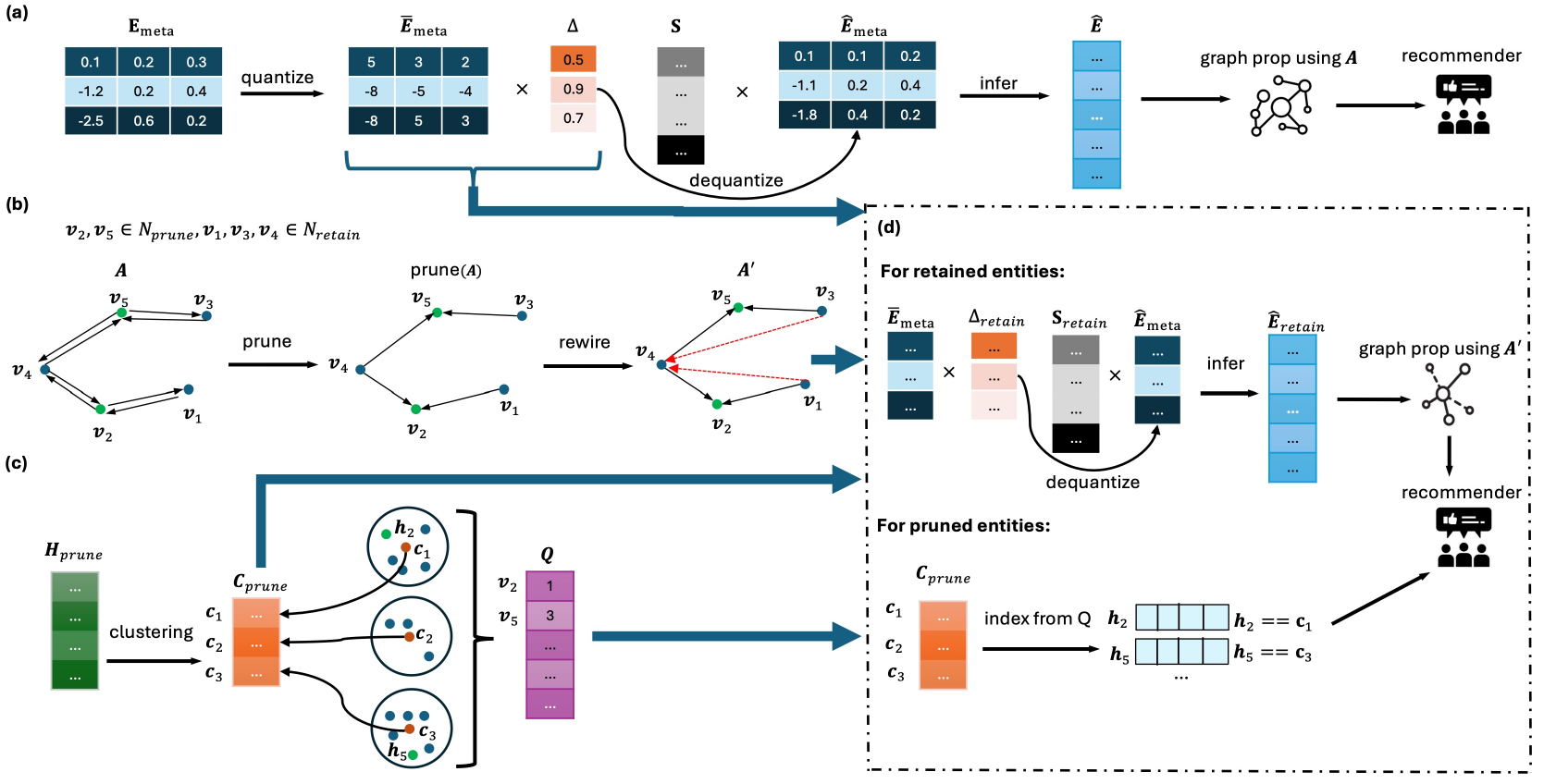}
    \caption{The overall workflow of \workname. (a) corresponds to the quantized compositional embedding table pretraining stage described in Sec. \ref{sec:pretrain}, (b) corresponds to graph rewiring for graph sparsification in Sec. \ref{sec:graph_rewiring}, (c) corresponds to pruned entity embedding imputation in Sec. \ref{sec:finetuning}, (d) corresponds to entity embeddings generation in the inference and fine-tuning stages described in Sec. \ref{sec:finetuning}. (a), (b), (c) are all conducted on the resource-rich server side, the fine-tuning stage in (d) can be conducted either on-server or on-edge. The red dotted arrows in (b) indicate the constructed edges between $v_4$ and her indirect neighbors after the graph rewiring process. 
    }
    \label{fig:overview}
\end{figure}

In this section, we detail the components of \workname, which includes a quantized compositional embedding table and a rewired propagation graph for embedding fine-tuning on edge devices. For convenience, we summarize the list of important symbols and notations introduced in this section in Tab. \ref{tab:symbol}. We also attach the framework visualization diagram in Fig. \ref{fig:overview}, which includes $4$ important stages: (a) quantized compositional embedding table pretraining; (b) graph rewiring for graph sparsification; (c) placeholder meta-embedding generation and assignment for pruned entities and; (d) entity embedding generation on edge devices for inference (and fine-tuning).

\begin{table}[h]
\caption{Table of symbols and notations.}
\label{tab:symbol}
\centering
\resizebox{\textwidth}{!}{
\begin{tabular}{ll}
\toprule
Symbol                               & Explanation                                                                                        \\ \hline
$\mathcal{U}$                        & Set of users.                                                                                      \\
$\mathcal{I}$                        & Set of items.                                                                                      \\
N                                    & Total number of users and items.                                                                   \\
$\textbf{R}$                         & Binarized user-item interaction matrix.                                                            \\
$\textbf{A}$                         & Full user-item interaction graph.                                                              \\
$\textbf{E}_{meta}$                  & Original compositional codebook.                                                    \\
$c$                                  & Number of meta-embeddings in the compositional codebook.                            \\
$\textbf{S}$                         & Highly sparse assignment weight matrix.                                                            \\
$\Delta$                             & Learnable step size vector for quantization.                                                       \\
$\Bar{\textbf{E}}_{meta}$            & Quantized compositional codebook.                                                   \\
$b$                                  & Bit length for quantization.                                                                       \\
$\hat{\textbf{E}}_{meta}$            & De-quantized compositional codebook.                                                \\
$\hat{\textbf{E}}$                   & Inferred full embedding table from the quantized compositional embedding table.                    \\
$\Bar{\textbf{E}}_{meta}^{pretrain}$ & Pretrained quantized compositional codebook.                                        \\
$\Delta^{pretrain}$                  & Pretrained learnable step size vector.                                                                       \\
$\textbf{H}^{pretrain}$              & Pretrained graph-propagated full embedding table.                                                  \\
$L$                                  & Total number of graph propagation layers.                                                          \\
$\textbf{B}$ & Entity-entity similarity matrix.\\
$m$                                  & Number of entities to be retained in the rewired graph.                                            \\
$o$                                  & Predefined rounding boundary for the linear programming problem formulated for the graph rewiring process. \\
$T$                                  & The highest propagation degree used in the graph rewiring algorithm.\\
$\mathcal{N}_{retain}$               & Set of retained entities in the rewired graph.                                               \\
$\mathcal{N}_{prune}$                & Set of entities to be refrained from propagating their collaborative signals in the rewired graph. \\
$\textbf{A}^\prime$                  & Rewired propagation graph used in the embedding fine-tuning and inference stages.                                  \\
$\textbf{H}_{retain}$                & graph-propagated embeddings for retained entities in the fine-tuning stage.                         \\
$\textbf{C}_{prune}$                 & Placeholder meta-embeddings for pruned entities.                                           \\
$q$                                  & Number of placeholder meta-embeddings in $\textbf{C}_{prune}$.                                     \\
$Q$                                  & Placeholder assignment vector for pruned entities.                                                 \\
$\hat{\textbf{H}}_{prune}$           & Imputed embedding table for pruned entities.                                                        \\ \hline
\end{tabular}}
\end{table}

\subsection{Preliminaries} \label{sec:problem_def}
In this work, we focus on the common ID-based recommendation, wherein GNN-based recommender systems are a primary choice for interaction modeling \cite{he_lightgcn_2020, sun2021hgcf, wang_neural_2019, chen2020revisiting, mao2021ultragcn, yu2022graph, yu2023xsimgcl}. We denote the set of users $\mathcal{U}$ and the set of items $\mathcal{I}$. The total number of users and items is $N = |\mathcal{U}| + |\mathcal{I}|$ and each user/item is assigned a unique ID $j \in [1, N]$. The user-item interaction matrix can be defined as a binary matrix $\textbf{R} \in \{0, 1\}^{|\mathcal{U}| \times |\mathcal{I}|}$ with $0$ and $1$ respectively indicating unobserved and observed interactions. In GNN-based recommender systems, the signal propagation is often conducted via a graph adjacency matrix $\textbf{A} \in \mathbb{R}^{N \times N}$ created using the interaction matrix $\textbf{R}$:
\begin{equation} \label{eq:original_graph}
   \textbf{A} = \begin{bmatrix} \textbf{0}& \textbf{R}\\\textbf{R}^\top&\textbf{0} \end{bmatrix}.
\end{equation}

Conventional GNN-based recommender systems make use of a full embedding table $\textbf{E} \in \mathbb{R}^{N \times d}$, where each user/item is assigned a physically unique embedding vector $e_j \in \mathbb{R}^d$. To obtain graph-propagated embeddings $\textbf{H} \in \mathbb{R}^{N \times d}$, the full embedding table $\textbf{E}$ is treated as the input embedding $\textbf{H}^{(0)}$. The hidden embeddings of the next layer $\textbf{H}^{(l+1)}$ is generated by propagating the hidden embeddings of the current layer $\textbf{H}^{l}$ with the symmetrically normalized graph adjacency matrix $\textbf{D}^{-\frac{1}{2}} \textbf{A} \textbf{D}^{-\frac{1}{2}}$, where $l \in [1, L]$ is the current layer number, $L$ is the total number of propagation layers and $\textbf{D} \in \mathbb{R}^{N \times N}$ is the diagonal degree matrix of $\textbf{A}$. The final graph-propagated embeddings $\textbf{H} \in \mathbb{R}^{N \times d}$ can then be generated by taking the mean of all hidden embeddings. The above descriptions are formulated as:
\begin{equation} \label{eq:original_graph_prop}
    \begin{aligned}
        \textbf{H}^{(0)} &= \textbf{E},\\
        \textbf{H}^{(l+1)} &= (\textbf{D}^{-\frac{1}{2}} \textbf{A} \textbf{D}^{-\frac{1}{2}}) \textbf{H}^{(l)},\\
        \textbf{H} &= \frac{1}{L + 1} \sum^{L}_{l=0} \textbf{H}^{(l)}.
    \end{aligned}
\end{equation}
In downstream recommendation settings, for user $u \in \mathcal{U}$ and item $i \in \mathcal{I}$, their graph-propagated embeddings $\textbf{h}_u, \textbf{h}_i \in \mathbb{R}^d$ are drawn from $\textbf{H}$ to perform affinity score calculation:
\begin{equation}
    \hat{y}_{ui} = \textbf{h}_{u}^\top \textbf{h}_i. 
\end{equation}
The predicted score $\hat{y}_{ui}$ is then inputted to common recommendation loss functions, such as the Bayesian personalized ranking (BPR) loss  \cite{rendle_bpr_2009}, for embedding optimization: 
\begin{equation} \label{eq:bpr_loss}
    \mathcal{L}_{\textit{BPR}} = \sum_{(u, i^+, i^-) \in \mathcal{B}} - \ln{\sigma(\widehat{y}_{ui^+}  - \widehat{y}_{ui^-})} + \lambda ||\Theta||^2,
\end{equation}
where $\mathcal{B}$ is either the whole training set or a training batch, $(u, i^+, i^-)$ is a triplet that contains sampled user $u$'s observed interacted item $i^+$ and unvisited item $i^-$, $||\Theta||^2$ is the $L_2$ regularization over trainable parameters and $\lambda$ controls its weight in the loss.

\subsection{Compositional Embedding Table} \label{sec:compositional_embedding}
In Sec. \ref{sec:problem_def}, we mention that the full embedding table is widely deployed in GNN-based recommenders. However, amid the increasing scale of users and items in recommendation services, the storage cost of the embedding table also soars rapidly, introducing difficulties in hosting large-scale datasets on edge devices as they are typically equipped with limited storage space. 

To reduce the storage cost, we instead opt for a compositional embedding table, which contains far fewer entries than the original full embedding table $\textbf{E}$. We follow our previous work \cite{liang2024lightweight} to devise a dense codebook $\textbf{E}_{meta} \in \mathbb{R}^{c \times d}$ and a highly sparse assignment weight matrix $\textbf{S} \in \mathbb{R}^{N \times c}_{\geq 0}$, where $c$ is the number of meta-embeddings stored in the codebook $\textbf{E}_{meta}$. Each row of the assignment matrix $\textbf{S}$ corresponds to a user/item, indicating which of the meta-embeddings are used to compose this entity's embedding vector as well as their weights in the composition. 

\textbf{Assignment Matrix Initialization.} 
Given that similar entities are expected to own similar meta-embeddings, a good initialization of assignment matrix $\textbf{S}$ can potentially contribute to higher recommendation accuracy and a speedup in training.  
In $\textbf{S}$, for each entity $p$, we term its highest-weighted meta-embedding (among the $t$ selected) the \textit{anchor meta-embedding} indexed by $q^*_{p}$, and it forms the base image of the entity's embedding $\textbf{e}_p$ since it takes the largest proportion in $\textbf{e}_p$'s composition.
To let correlated entities have similar base embedding images during initialization, we propose to assign two entities $p_1$, $p_2$ the same anchor meta-embedding weight if they demonstrate a high affinity with each other, \ie $\textbf{S}[p_1, q^*_{p_1}]=\textbf{S}[p_1, q^*_{p_2}]$.
Because the entities lying within each community in fact form tightly connected subgraphs in the full interaction graph, assigning them the same anchor meta-embedding helps pass the same proximity onto the latent space. In our work, we take advantage of a well-established multilevel graph partitioning algorithm, namely METIS \cite{karypis1997metis} for partitioning the user-item interaction graph. Although other advanced graph clustering methods can also be considered, METIS well serves the one-off initialization purpose due to its fast and accurate computation (no learning involved), balanced and non-overlapping partitions (all anchor meta-embeddings are fairly utilized), as well as deterministic results (ease of replication). 
We set the desired partition number to $c$, where entities in the same subgraph share one specific anchor meta-embedding in $\textbf{E}_{meta}$. 
To reflect this, on initialization of the assignment matrix $\textbf{S}$, we perform the following for every entity $p$: 
\begin{itemize}[leftmargin=*]
\item[I.] Given entity $p$'s subgraph index $c_p\in \{1,2,...,c\}$, we set $\textbf{S}[p,q_p^*]= w^*$, where $q_p^*=c_p$ and $w^*$ is a universal hyperparameter.
\item[II.] We uniformly sample $t-1$ indexes from $\{1,2,...,c\}\setminus c_p$ with replacement, denoted as a set $\mathcal{Q}$. For every $q\in \mathcal{Q}$, we set the corresponding assignment weight $\textbf{S}[p,q]=\frac{(1 - w^*)}{t-1}$. 
\item[III.] For all remaining entries at $q'\notin \mathcal{Q}\cup c_p$, we set $\textbf{S}[p,q^{\prime}]=0$.
\end{itemize}
In short, each entity receives an initial anchor meta-embedding based on the subgraph assigned, while the remaining $t-1$ meta-embedding assignments are randomly initialized in favor of diversity. Our hyperparameter study in \cite{liang2024lightweight} shows that so long as there is more than one meta-embedding assigned to each user/item with the anchor meta-embedding $q_p^{*}$ carefully selected on initialization, a competitive recommendation performance is guaranteed. Hence, we select $2$ meta-embeddings for each user/item, one of which is the anchor meta-embedding $q_p^{*}$ selected by following step I of the sampling strategy mentioned above. The other auxiliary meta-embedding $q$ is randomly drawn from the codebook as depicted in step II above to encourage the uniqueness of composed entity embeddings. Following step III, all remaining entries in $\textbf{S}$ have the value of $0$. In this way, we ensure the assignment matrix $\textbf{S}$ is highly sparse, as each row has only $2$ nonzero elements. In our previous work 
\cite{liang2024lightweight}, we study the impact of different weights assigned to the anchor meta-embedding $w^*$. Our finding is that setting $w^* = 0.9$ yields a satisfactory recommendation performance when the dataset contains a relatively large number of users and items. Therefore, 
\workname \text{} follows the weight distribution scheme of 0.9/0.1 split between the anchor meta-embedding $q_p^*$ and auxiliary meta-embedding $q$ as well. 

To infer the full embedding table $\hat{\textbf{E}} \in \mathbb{R}^{N \times d}$  for downstream recommendation tasks, one can simply perform the matrix multiplication between $\textbf{S}$ and $\textbf{E}_{meta}$:
\begin{equation} \label{eq:impute_full_emb}
    \hat{\mathbf{E}} = \textbf{S} \textbf{E}_{meta}. 
\end{equation}

With the assignment matrix $\textbf{S}$, one can also flexibly update composition weights assigned to each meta-embedding for each user/item to reflect the change of tendency toward meta-embeddings over time. For example, in LEGCF \cite{liang2024lightweight}, we design an expanded interaction graph that treats meta-embeddings as virtual nodes to perform simultaneous graph propagation between meta-embeddings and entity embeddings. The graph-propagated embeddings are then used to update the assignment weight matrix $\textbf{S}$ by solving a similarity constraint. 

By replacing the full embedding table $\textbf{E}$ with a dense codebook $\textbf{E}_{meta}$ and a highly sparse assignment weight matrix $\textbf{S}$, the storage cost of the embedding layer can be greatly reduced from $O(Nd)$ to $O(cd + 2N)$. The compositional embedding structure performs row-wise compression on the embedding table, since the number of unique embedding vectors to be stored on the disk is lowered.

\subsection{Codebook Quantization-Aware Training} \label{sec:qat_training}
In Sec. \ref{sec:compositional_embedding}, we have detailed the core components and working mechanisms of our prior work LEGCF. From this section onward, we unfold the design of its enhanced version, namely \workname. 

In LEGCF, despite that we leverage the codebook to bring down the number of embedding rows to be stored on disk, each unique embedding vector is still saved as a sequence of floating point numbers. Since the size of each floating point number is $4$ bytes, to store a single meta-embedding with a relatively large dimension size, say 128, $512$ bytes will be utilized. For edge devices with limited storage space, the number of usable meta-embeddings can be throttled by the size of a single meta-embedding vector. To solve this, there has been a broad coverage of research focusing on compressing elements in embeddings or model parameters using quantization \cite{nagel2021white, yao2021hawq, wang2023bitnet,liullm}. The key idea is to map each continuous floating point number element in the embedding vector into discrete integer chunks so that the elements can be stored efficiently using low-bit fixed-point representations, like INT8/16. As such, the storage size of each embedding element can be squeezed to 1 byte and 2 bytes respectively, reducing the original storage size effectively by $75\%$ and $50\%$. To learn accurate fixed-point representations, quantization-aware training (QAT) has been a popular training paradigm for quantizing the model parameters \cite{liullm, du2024bitdistiller, wang2023bitnet, chen2024efficientqat, ma2024era} due to its compatibility with gradient descent update and ease of implementation \cite{yin2024device}. In QAT, a floating point precision embedding table is used as a backbone parameter for back-propagation updates. In forward passes, the full-precision embeddings are quantized into low-bit integer precision first and then de-quantized to floating point precision by multiplying a step size. In backward passes, the gradients of the full-precision embeddings are approximated by the straight through estimator (STE) \cite{bengio2013estimating} to perform value update. Since the downstream recommendation task takes the de-quantized embeddings from their quantized counterparts, the optimization objective is aware of the quantization process and aims to learn a full-precision embedding table that encodes the quantization semantics. 

In our framework, we adopt the learned step size quantization strategy (LSQ) \cite{esserlearned} on the dense codebook $\textbf{E}_{meta}$ to further perform element-wise compression. The low-bit integer representation of the codebook $\Bar{\textbf{E}}_{meta} \in \{-2^{b-1},.., 2^{b - 1} - 1\}^{c \times d}$ can be obtained as follows:
\begin{equation} \label{eq:quantized_E_meta}
    \Bar{\textbf{E}}_{meta} = \textit{Round}(\textit{Clip}(\frac{\textbf{E}_{meta}}{\Delta}, Q_{min}, Q_{max})),
\end{equation}
where $\textit{Round}(\cdot)$ is the integer rounding function, $\textit{Clip}(\textbf{x}, \textit{min}, \textit{max})$ ensures values in $\textbf{x}$ that are lower than \textit{min} are set to \textit{min} and values larger than \textit{max} are set to \textit{max}; $Q_{min} = -2^{b-1}, Q_{max} = 2^{b - 1} - 1$ represents the lowest and largest integer values that can be used to represent floating point numbers, $b$ is the bit length of the fixed-point representations, $\Delta \in \mathbb{R}^c$ is the learnable step size vector. The de-quantized codebook $\hat{\textbf{E}}_{meta} \in \mathbb{R}^{c \times d}$ is computed by multiplying $\Bar{\textbf{E}}_{meta}$ by the learnable step size vector $\Delta$ as:
\begin{equation} \label{eq:dequantize_E_meta}
    \hat{\textbf{E}}_{meta} = \Bar{\textbf{E}}_{meta} \times \Delta,
\end{equation}
one can then substitute $\textbf{E}_{meta}$ with $\hat{\textbf{E}}_{meta}$ in Eq. \ref{eq:impute_full_emb} to infer the full embedding table.

Since the dense codebook can now be stored as low-bit integer representations $\Bar{\textbf{E}}_{meta}$ and a floating point precision step size vector $\Delta$, the storage size of the codebook is reduced from $O(4\textit{ bytes} \times \text{ } cd)$ to $O(4\textit{ bytes } \times \text{ } c(\frac{b}{32}d+ 1))$. This implies that under a fixed storage budget, replacing the full-precision codebook $\textbf{E}_{meta}$ with our low-bit quantized codebook $\Bar{\textbf{E}}_{meta}$ allows more meta-embeddings to be included in the codebook. Thus, enriching the diversity of compositional embeddings and reducing the risk of entity embedding collisions. 

As most modern hardware has native support for INT8/INT16 data type \footnote{\url{https://www.intel.com/content/www/us/en/developer/articles/technical/int8-quantization-for-x86-cpu-in-pytorch.html}; \url{https://docs.qualcomm.com/bundle/publicresource/topics/80-63442-50/quantization.html}; \url{https://docs.nvidia.com/nemo-framework/user-guide/latest/nemotoolkit/nlp/quantization.html}.}, the quantized codebook $\Bar{\textbf{E}}_{meta}$ can be easily stored on edge devices without any external libraries or dedicated hardware.

\subsection{Codebook Pretraining} \label{sec:pretrain}
Due to the hardware limitation of edge devices and the desiderata of fast adaptation of GNN-based recommender systems that satisfy various hardware specifications on the edge side, it is strongly motivated to first conduct a one-off pretraining stage on the server side. In this way, only the transfer of pretrained framework components and a lightweight refinement process is required on the resource-constrained edge side to quickly adapt the recommender system. 

We propose to pretrain the quantized compositional embedding table. In each training batch, we combine the quantization-aware training process defined in Sec. \ref{sec:qat_training} with the graph propagation operation conducted on the inferred full embedding table $\hat{\textbf{E}} \in \mathbb{R}^{N \times d}$ using the user-item interaction graph $\textbf{A}$. This is done by replacing the input embeddings $\textbf{H}^{(0)}$ in Eq. \ref{eq:original_graph_prop} to the full embedding table inferred from the quantized compositional codebook:
\begin{equation} 
        \textbf{H}^{(0)} = \hat{\textbf{E}} = \textbf{S}\hat{\textbf{E}}_{meta} = \textbf{S}(\Bar{\textbf{E}}_{meta} \times \Delta).
\end{equation}
Despite in LEGCF \cite{liang2024lightweight}, we propose an assignment update strategy on the assignment weight matrix \textbf{S}, our ablation studies \cite{liang2024lightweight} find out that the initialization of \textbf{S} via the METIS graph partitioning algorithm \cite{karypis1997metis} is a dominating factor for the performance boost compared with performing additional updates on \textbf{S}. Moreover, the update strategy involves the graph propagation between an expanded interaction graph and a larger embedding table, which introduces an even higher runtime computation cost than the original full embedding table setting. For these reasons, in \workname, we no longer update the assignment weight matrix \textbf{S} after it has been initialized.

Once the learning of the quantized codebook and learnable step size vector converges, the pretraining stage is complete. We obtain the pretrained quantized codebook $\Bar{\textbf{E}}_{meta}^{pretrain}$ and pretrained step size vector $\Delta^{pretrain}$. We also save the pretrained graph-propagated full embedding table $\textbf{H}^{pretrain}$ and use it to perform graph rewiring in Sec. \ref{sec:graph_rewiring}.

\subsection{Graph Rewiring for Graph Sparsification} \label{sec:graph_rewiring}
For on-device deployment of GNN-based recommenders, the computational efficiency and runtime memory consumption can also be bottlenecks for large-scale recommendation scenarios. Due to the varying importance of nodes in the message passing process within a graph \cite{gao2023survey}, we point out that performing graph propagation using the full user-item interaction graph on edge devices introduces difficulty in deployment and thus, should be avoided. 

To alleviate the graph computational complexity problem, one effective solution is graph rewiring \cite{toppingunderstanding}, which modifies the structure of the graph by removing unnecessary nodes (\ie entities) and edges (\ie interactions). 
Our objective here for graph rewiring is to identify and remove less prominent propagation links from the user-item interaction graph $\textbf{A}$ so that the sparsity of the graph is increased to lower required MACs for graph propagation. To do this, it is intuitive to first identify the impact of each user/item in participating in collaborative signal propagation. Those that contribute little to no impact can be refrained from propagating their embeddings to their neighbors without causing a significant drop in recommendation performance. It is worth noting that many proposed GNN computation cost improvement work that utilizes the sampling strategy \cite{hamilton2017inductive, gupta2024graphscale,zeng2020graphsaint, chiang2019cluster, qu2023tt} can also be regarded as a form of graph rewiring, as the unsampled nodes and/or edges are turned off temporarily for graph propagation. However, these methods either require a complicated search process to identify an optimal subgraph, or a time-consuming node/edge sampling procedure is often introduced. There are also existing graph sparsification methods available \cite{rong2019dropedge, yang2019topology, chen2021unified, zheng2020robust}, but most of them still involve complicated optimization objectives that are computationally intensive, or require handcrafted sparsification rules that suffer from limited generalizability.  We argue that involving such a cumbersome and lengthy graph rewiring/sparsification strategy in our framework increases the difficulty in quick deployment, and instead look for a more time- and computation-efficient alternative. 

To determine which nodes/edges are to be dropped from the full interaction graph, the simplest way is to sort the nodes/edges by their degrees and remove the ones that provide limited connectivity within the graph. However, such a \naive solution overlooks the node affinity in the latent space, which may yield an ill-defined rewired graph and hurt the embedding quality when performing graph propagation.
Notice that from Sec. \ref{sec:pretrain}, we have the graph-propagated embedding table $\textbf{H}^{pretrain}$ pretrained using the full user-item interaction graph $\textbf{A}$, which conceals rich entity features and collaborative semantics.
We leverage the entity-entity similarity matrix $\textbf{\textit{B}} \in \mathbb{R}^{N \times N}$ computed based on the pretrained full embedding table $\textbf{H}^{pretrain}$ to determine entities' contribution in graph collaborative signal propagation:
\begin{equation} \label{eq:sim_mat}
    \textbf{\textit{B}} = \textbf{H}^{pretrain} \textbf{H}^{pretrain \top}.
\end{equation}
Each row of $\textbf{\textit{B}}$ corresponds to an entity $j$ and stores the similarity scores between this entity and all entities in the dataset. We denote each row as $\textbf{\textit{B}}_{j} \in \mathbb{R}^{N}$ so $\textbf{\textit{B}}_{jk}$ indicates the similarity score between an entity pair $(j,k)$. With the similarity scores calculated from Eq. \ref{eq:sim_mat}, we can now interpret the high-contribution entity selection task as a binary integer programming (BIP) problem with the objective of maximizing the dataset's overall entity-entity similarity scores:
\begin{equation} \label{eq:BIP}
\begin{aligned}
\textit{maximize}              &\; \sum_{j = 1}^{N} v_j (\sum_{k = 1}^{N}\textbf{\textit{B}}_{jk} ), \\
\textit{s.t.} &\;
\begin{alignedat}[t]{3}
  \sum_{j = 1}^{N} v_j &= m, \\
v_j &\in \{0, 1\}, j \in \{1, .., N\}, \\
\end{alignedat}
\end{aligned}
\end{equation}
where $v_j$ is a binary variable that corresponds to the entity $j$, such that entities having little to no contribution to signal propagation are assigned a value of $0$. The ones with higher impact have a value of $1$ instead. $m \in \{1, .., N\}$ is a hyperparameter that controls the number of entities to be retained after the graph rewiring process. 
By setting different values of $m$, one can accommodate various graph computation and memory efficiency requirements. For convenience, we define $m$ to be proportional to the total number of users and items in the dataset, and we name this ratio the \textbf{retention ratio}. For example, if the retention ratio is set to $0.7$ for a dataset, the value of $m$ is calculated as $m = \lfloor 0.7N \rfloor$.
Our intention of formulating a BIP problem on the entity-entity similarity matrix constructed using graph pretrained embeddings to realize the impact of entities is intuitive: given that at most $m$ nodes can remain in the rewired graph, we aim to find the ones such that the sum of the similarity scores in the rewired graph is maximized. Since graph propagation encodes neighbor collaborative signals in entity embeddings \cite{wang_neural_2019, he_lightgcn_2020}, an entity that is similar to a large number of entities in the graph hidden space implies her significance in graph propagation and strong connectivity in the user-item interaction graph. Thus, retaining the set of such entities in the rewired graph maximally preserves the overall collaborative semantics in the user-item interaction graph. On the contrary, entities with embeddings different from the majority are usually less connected or exhibit high heterophily in the user-item interaction graph. These entities are deemed less impactful on graph propagation. Hence, we prevent them from participating in collaborative signal propagation in the rewired graph to lower graph computation complexity.
Despite the simplicity of Eq. \ref{eq:BIP}, solving BIP directly has been proven to be NP-complete \cite{karp2010reducibility}, which means there is no guarantee that a valid solution can be found in polynomial time, challenging the computational efficiency. To ensure a low computational overhead for graph rewiring, we propose to first relax the BIP into a linear programming (LP) problem by substituting binary variables $v_j, j \in \{1, .., N\}$ with continuous numbers $\hat{v}_j, j \in \{1, .., N\}$ in the $[0, 1]$ range:
\begin{equation} \label{eq:LP}
\begin{aligned}
\textit{maximize}              &\; \sum_{j = 1}^{N} \hat{v}_j (\sum_{k = 1}^{N}\textbf{\textit{B}}_{jk} ), \\
\textit{s.t.} &\;
\begin{alignedat}[t]{3}
  \sum_{j = 1}^{N} \hat{v}_j &= m, \\
\hat{v}_j &\in [0, 1], j \in \{1, .., N\}, \\
\end{alignedat}
\end{aligned}
\end{equation}

With this relaxation, the problem can be solved using the simplex algorithm \cite{dantzig1990origins}, which yields polynomial-time complexity \cite{schrijver1998theory}. We then approximate the binary variables by applying value rounding with a predefined boundary $o 
\in [0, 1]$:
\begin{equation} \label{eq:approx_v_i}
    v_j = 
    \begin{cases}
        1 & \text{, if } \hat{v}_j > o, \\
        0 & \text{, otherwise}. 
    \end{cases}
\end{equation}
Taking the set of binary variables $v_j, j \in \{1, .., N\}$ gathered from Eq. \ref{eq:approx_v_i}, we define the set of entities with high impact on graph propagation as $\mathcal{N}_{retain}$ (\ie \textbf{retained entities}), the set of entities with relatively lower contribution on graph propagation as $\mathcal{N}_{prune}$ (\ie \textbf{pruned entities}) and $| \mathcal{N}_{retain}| = m, |\mathcal{N}_{prune}| = N - m, \mathcal{N}_{retain} \cap \mathcal{N}_{prune} = \emptyset$. Given an undirected graph $\textbf{A} \in \mathbb{R}^{N \times N}$ represented in the form of an adjacency matrix, we define a graph sparsification function that increases the graph sparsity by nullifying the columns with indices corresponding to the IDs of the pruned entities:
\begin{equation} \label{eq:prune_func}
    \textit{prune}(\textbf{A}_{:,j}) = \begin{cases}
        0 & \text{, if } j \in \mathcal{N}_{prune}, \\
        \textbf{A}_{:, j} & \text{, otherwise},
    \end{cases}
\end{equation}
where $\textbf{A}_{:,j}$ is the $j^{th}$ column of the full user-item interaction graph.
In this way, the sparsified graph becomes a directed graph, where the entities in $\mathcal{N}_{prune}$ no longer propagate their collaborative signals to their neighbors, yet they can still receive signals from their neighbors to update embeddings, given their corresponding rows are not all zeros. If we apply Eq. \ref{eq:prune_func} to the full interaction graph $\textbf{A}$, however, the resultant sparsified graph may introduce additional all zero rows. The entities with all zero rows mean that the set of first-hop neighbors is a subset of the pruned entity set $\mathcal{N}_{prune}$, so there are no neighbors to receive the collaborative signals from. 

\begin{algorithm}
    \caption{Graph rewiring algorithm.}
    \label{alg:graph_rewire}
    \textbf{Input}: full user-item interaction graph $\textbf{A}$, total number of entities $N$, the highest propagation degree $T$ and multi-hop adjacency matrices $A^{1},..,A^T$ \\
     \textbf{Output}: the sparsified rewired propagation graph $\textbf{A}^\prime$
     
       $\textbf{A}^\prime \gets \textit{prune}(\textbf{A})$\\
        \For{$j \in [1, .., N]$}{
            \If{ $\textit{count\_nonzero}(\textbf{A}^\prime_j) = 0$}{
                \For{$t \in [2, .., T]$}{
                    $g^\prime \gets \textit{prune}(\textbf{A}^t_j)$\\
                    \If{$\textit{count\_nonzero}(g^\prime) > 0$}{
                        $\textbf{A}^\prime_j \gets \textit{sign}(g^\prime)$\\
                        \textit{break}
                    }
                }
            }
        }
\end{algorithm}

We argue that it is crucial to have some neighbors for each entity so that their embeddings can be improved during the graph propagation step. To this end, we construct a rewired propagation graph $\textbf{A}^\prime \in \mathbb{R}^{N \times N}$ by taking the row-wise disjoint of column nullified interaction graphs of various propagation degrees. This process is displayed in Alg. \ref{alg:graph_rewire},
where $T$ is the highest propagation degree used, $\textit{count\_nonzero}(\cdot)$ returns the number of nonzero elements in a vector, $\textit{sign}(\cdot)$ is the signum function. Simply put, line 1 nullifies pruned entities' corresponding columns in the first-hop neighborhood. After that, we find all entities that do not have first-hop neighbors left and fill their receptive fields with their second-hop neighbors or third-hop neighbors given that they are not members in $\mathcal{N}_{prune}$, so on and so forth until the rows contain some nonzero values (lines 4 - 8). The signum function in line 7 ensures all indirect neighbors have the same weight as the first-hop neighbors. Following this, the number of all zero rows caused by the column nullification process can be greatly reduced, while the overall sparsity of the interaction graph increases, which lowers the required MACs when performing graph propagation. In practice, setting $T = 4$ in most cases is sufficient to entirely recover all zero rows caused by the nullification process by rewiring these entities with their indirect neighbors. Since Alg. \ref{alg:graph_rewire} utilizes various multi-hop adjacency matrices, they are precomputed and stored on the resource-rich server side so that one can reuse them for multiple graph rewiring procedures without incurring noticeable counter effects on framework efficiency. 

The rewired propagation graph $\textbf{A}^\prime$ will be transmitted to resource-constrained edge devices for inference. To meet various hardware capacities of edge devices, one can adjust the number of retained entities $m$ to modify the message passing complexity of the rewired graph and achieve a desired level of efficiency-accuracy trade-off. To ensure the optimality of the quantized compositional embedding table, we propose conducting a quick fine-tuning process using the rewired propagation graph $\textbf{A}^{\prime}$. We detail the embedding fine-tuning process in Sec. \ref{sec:finetuning}.

\subsection{Fine-tuning with Rewired Graph} \label{sec:finetuning}
With the pretrained quantized codebook $\Bar{\textbf{E}}_{meta}^{pretrain}$, the pretrained learnable step size vector $\Delta^{pretrain}$ and the rewired propagation graph $\textbf{A}^\prime$ obtained from Sec. \ref{sec:pretrain} and Sec. \ref{sec:graph_rewiring}, we can perform a light embedding fine-tuning process to fit the embedding table to various storage and memory constraints. We take the inferred embeddings $\hat{\textbf{E}}_{retain} \in \mathbb{R}^{m \times d}$ of the retained entities as the input embeddings and perform propagation over the newly generated rewired graph $\textbf{A}^\prime$ to conduct quantization-aware training: 
\begin{equation} \label{eq:refine_graph_prop}
    \begin{aligned}
        \textbf{H}_{retain}^{(0)} &= \hat{\textbf{E}}_{retain} = \textbf{S}_{retain} \hat{\textbf{E}}_{meta} = \textbf{S}_{retain}(\Bar{\textbf{E}}_{meta} \times \Delta),\\
        \textbf{H}_{retain}^{(l+1)} &= (\textbf{D}^{-\frac{1}{2}} \textbf{A}^\prime \textbf{D}^{-\frac{1}{2}}) \textbf{H}^{(l)}_{retain},\\
        \textbf{H}_{retain} &= \frac{1}{L + 1} \sum^{L}_{l=0} \textbf{H}_{retain}^{(l)},
    \end{aligned}
\end{equation}
where $\textbf{S}_{retain} = \textbf{S} [\mathcal{N}_{retain}, :] \in \mathbb{R}^{m \times c}$ is the assignment weight matrix for entities in $\mathcal{N}_{retain}$; $\Delta \in \mathbb{R}^{c}$ is the learnable step size vector; $\textbf{D} \in \mathbb{R}^{m \times m}$ is the diagonal degree matrix of $\textbf{A}^\prime$. On initialization, we set $\Bar{\textbf{E}}_{meta}$ to be $\Bar{\textbf{E}}_{meta}^{pretrain}$ and $\Delta$ to be $\Delta^{pretrain}$. Essentially, in the fine-tuning step, only the embeddings of retained entities will be updated using the rewired graph $\textbf{A}^\prime$. The advantages of this embedding refinement procedure are three-fold. Firstly, updating only the retained entities' embeddings caps the size of the computable matrices at forward and backward passes at $O(md)$, which effectively controls the peak runtime memory of \workname \textbf{} at training and inference. The use of the rewired graph, on the other hand, lowers the total MACs in graph propagation. Thus, the issue with the expensive graph computational cost on edge devices is alleviated. The two innovations together enable the smooth deployment and fine-tuning (if necessary) of a GNN-based recommender system directly on-edge. Secondly, a fast and lightweight embedding fine-tuning strategy boosts the time efficiency for large-scale edge device deployment, as most embedding optimization work requires a complete training process for each hardware specification. In \workname, instead, for each hardware configuration, one only needs to input a suitable retained entity budget $m$ and fine-tune the quantized compositional embedding table using the generated rewired graph to obtain a lightweight GNN-based recommender with satisfactory accuracy. 
Lastly, refining the quantized codebook and learnable step size vector using only retained entities' graph-propagated embeddings encodes the entity importance in the quantized codebook as now the quantized codebook is driven to improve the retained entities' embeddings, of who is deemed impactful in graph propagation.

Since only the embeddings of retained entities are inferred from the quantized codebook and propagated through the rewired graph, the graph-propagated embeddings of pruned entities need to be imputed using other strategies. Despite one can directly draw pruned entities' graph-propagated embeddings $\textbf{H}^{pretrain}_{prune} = \textbf{H}^{pretrain}[\mathcal{N}_{prune}, :] \in \mathbb{R}^{(N-m) \times d}$ from the pretrained full embedding table $\textbf{H}^{pretrain}$ and store it on edge devices for inference, this \naive solution contradicts with the concern on embedding table storage cost.
As a wraparound, we design a placeholder codebook $\textbf{C}_{prune} \in \mathbb{R}^{r \times d}$ generated by applying the clustering technique such as the K-means algorithm \cite{lloyd1982least} on $\textbf{H}_{prune}^{pretrain}$ for pruned entity embeddings imputation:
\begin{equation} \label{eq:placeholder_emb}
        \textbf{C}_{prune}, Q = \textit{clustering}(\textbf{H}_{prune}^{pretrain}, r),
\end{equation}
where $Q \in [1, .., r]^{N-m}$ is the placeholder assignment vector for entities in $\mathcal{N}_{prune}$, $r$ is the number of placeholder meta-embeddings in the codebook $\textbf{C}_{prune}$. To save the storage space, we have $r \ll (N - m)$. By letting pruned entities share a smaller pool of placeholder embeddings, one only needs to store $\textbf{C}_{prune}$ and $Q$ on disk. Assuming $Q$ has the precision of INT32, the space complexity of pruned entity embedding is $O(4\textit{ bytes } \times \text{ } (rd + (N - m))$. It is reasonable to set $r \ll c$ since the pruned entities have less impact in graph propagation than the retained entities, whose embeddings are inferred from the quantized codebook $\Bar{\textbf{E}}_{meta}$ and the rewired propagation graph $\textbf{A}^\prime$.
At inference time, the imputed embedding table of pruned entities $\hat{\textbf{H}}_{prune} \in \mathbb{R}^{(N-m) \times d}$ is created by taking the placeholder embeddings with respect to the assignment vector:
\begin{equation} \label{eq:get_pruned_embs}
    \hat{\textbf{H}}_{prune} = \textbf{C}_{prune}[Q].
\end{equation}

\begin{algorithm}
     \textbf{Input}: pretrained quantized codebook $\Bar{\textbf{E}}_{meta}^{pretrain}$ and pretrained step size vector $\Delta^{pretrain}$, rewired propagation graph $\textbf{A}^\prime$, placeholder codebook $\textbf{C}_{prune}$ and placeholder assignment vector $Q$ \\
     \textbf{Output}: fine-tuned quantized codebook $\Bar{\textbf{E}}^{*}_{meta}$ and fine-tuned step size vector $\Delta^{*}$
    \caption{Fine-tuning process.}\label{alg:finetune}
        Initialize $\Bar{\textbf{E}}_{meta}$ as $\Bar{\textbf{E}}_{meta}^{pretrain}$

        Initialize $\Delta$ as $\Delta^{pretrain}$
        
        \While{not converged}{
        Follow Eq. \ref{eq:refine_graph_prop} to refine quantized codebook $\Bar{\textbf{E}}_{meta}$ and step size vector $\Delta$ using retained entity embeddings and rewired propagation graph $\textbf{A}^\prime$

        Draw pruned entities' embeddings $\Hat{\textbf{H}}_{prune}$ w.r.t. Eq. \ref{eq:get_pruned_embs} for inference
        
        }
       
\end{algorithm}
The fine-tuning process is depicted in Alg. \ref{alg:finetune}. Note that the placeholder meta-embedding generation for pruned entities is not part of the fine-tuning process as we refine the quantized compositional embedding table only using the retained entity embeddings. Due to the use of a rewired graph and only a part of the entity embeddings participate in graph propagation, the fine-tuning process incurs low runtime memory consumption and hence, can be conducted either on-sever or on-edge efficiently.

\subsection{Framework Overview}
\begin{algorithm}
    \caption{\workname \text{} implementation.}\label{alg:overall}
        Initialize $\textbf{S} \in \mathbb{R}^{N \times c}$ as defined in Sec. \ref{sec:compositional_embedding} 
        
        Formulate codebook quantization-aware training as stated in Sec. \ref{sec:qat_training}

        \While{not converged}{
        Pretrain quantized codebook $\Bar{\textbf{E}}_{meta}$ and learnable step size vector $\Delta$ by following Sec. \ref{sec:pretrain}
        }

        Follow Sec. \ref{sec:graph_rewiring} and Alg. \ref{alg:graph_rewire} to generate rewired propagation graph $\textbf{A}^\prime$ using the pretrained full embedding table $\textbf{H}^{pretrain}$

        Generate placeholder meta-embeddings $\textbf{C}_{prune}$ and placeholder assignment vector $Q$ for pruned entities w.r.t. Eq. \ref{eq:placeholder_emb} 

        Follow Alg. \ref{alg:finetune} to fine-tune the quantized compositional embedding table using the rewired graph.

\end{algorithm}

To illustrate the implementation of \workname, we provide the overall algorithm in Alg. \ref{alg:overall}. Lines 3-4 conduct quantized compositional codebook pretraining. Line 5 performs the graph rewiring step to obtain a sparsified propagation graph to be used on edge devices. Line 6 generates the placeholder meta-embeddings and assigns them to pruned entities. Line 7 carries out the fine-tuning procedure on the pretrained quantized compositional codebook. Lines 3-6 are executed on the resource-rich server due to their relatively high computation cost and runtime memory consumption. Line 7 has the option to be executed on-edge or on-server. 

To analyze the space complexity of \workname, the quantized compositional codebook $\Bar{\textbf{E}}_{meta}$ and the floating point precision learnable step size vector $\Delta$ exert space of $O(4 \textit{ bytes } \times \text{ } c(\frac{b}{32}d+ 1))$, the floating point placeholder meta-codebook $\textbf{C}_{prune}$ along with the INT32 placeholder assignment vector $Q$ incur space of $O(4 \textit{ bytes } \times \text{ } (rd + (N - m))$, so the total space complexity is $O(c(\frac{b}{32}d+ 1)) + (rd + (N - m)))$, which is far lower than the space complexity of a full embedding table $O(Nd)$. Since commonly $r \ll c$, the former term has higher significance when evaluating the overall space efficiency.

\section{Experiments} \label{sec:experiments}
In this section, we carry out experiments to verify the effectiveness of \workname. This section is organized to answer the following research questions (RQs):
\begin{itemize}
    \item \textbf{RQ1:} Does our framework work well compared to other baselines under resource-constrained settings?
    \item \textbf{RQ2:} What is the effect of important components proposed in \workname?
    \item \textbf{RQ3:} How do different hyperparameter settings affect our framework's performance?
    \item \textbf{RQ4:} Does \workname \text{} work well when equipped with other GNN-based  recommenders?
    \item \textbf{RQ5:} How does the precision of the quantized codebook affect recommendation performance?
    \item \textbf{RQ6:} How efficient is our proposed on-device fine-tuning process?
\end{itemize}

\subsection{Experimental Settings}

\subsubsection{Datasets}\label{sec_dataset}
We conduct experiments on three publicly available datasets: two medium-scale datasets \textbf{Yelp2020} and \textbf{Amazon-book} as well as one industrial large-scale dataset \textbf{Alibaba-iFashion} (iFashion). Yelp2020 dataset is downloaded from \cite{sun2021hgcf}, Amazon-book dataset can be found in \cite{he_lightgcn_2020} and iFashion dataset is readily available from \cite{chen2019pog}. For Yelp2020 and Amazon-book, we use all user-item interactions in the dataset. For iFashion dataset, due to the high sparsity of observed interactions, we follow the practices detailed in \cite{wu2021self} to sample $500,000$ users and include all user-item interactions made by the sample users to form the dataset for experiment. The detailed statistics of the three datasets are depicted in Tab. \ref{tab:dataset_stat}. We adopt the train/test/validation split ratios of 80\%/20\%/10\% on observed interactions. For each observed user-item interaction in the training set, we sample 5 negative items and add them to the training set.

\begin{table}[h]
    \caption{Statistics of datasets used in experiments.}
    \label{tab:dataset_stat}
    \centering

    \begin{tabular}{c  c c  c c}
        \toprule
        Dataset & \#User & \#Item & \#Interaction & Density\\
        \hline
        Yelp2020  & 71,135  & 45,063 & 1,782,999 & 0.056\%\\
        Amazon-book & 52,643 & 91,599 & 2,984,108 & 0.062\%\\
        iFashion & 500,000 & 1,465,817 & 26,069,309 & 0.004\%\\
        \bottomrule
    \end{tabular}
\end{table}

\subsubsection{Based Recommenders and Baselines}
Since \workname \text{} is a lightweight framework dedicated to GNN-based recommender systems, we select LightGCN \cite{he_lightgcn_2020} as the base recommender for main experiments considering its effectiveness and popularity, and we also verify our framework's generalization with other GNN-based recommender in Section~\ref{sec_rq4}. 

To show the effectiveness of our framework, we compare the performance of \workname \text{} against a wide range of embedding optimization state-of-the-art baselines:
\begin{itemize}
    \item \textbf{ESAPN} \cite{liu_automated_2020}: An AutoML-based dimension search algorithm that devises reinforcement learning to find the suitable user and item embedding sizes that yield an optimal recommendation performance.
     \item \textbf{CIESS} \cite{qu2023continuous}: The state-of-the-art AutoML-based lightweight embedding framework which allows user's input of desirable storage budget.
    \item \textbf{PEP} \cite{liu_learnable_2021}: The iconic pruning-based strategy which identifies and nullifies unimportant elements in the full embedding table.
    \item \textbf{OptEmbed} \cite{lyu_optembed_2022}: An embedding optimization work that combines both pruning and evolutionary search.
   \item \textbf{CERP} \cite{liang2023learning}: The state-of-the-art pruning method which conducts simultaneous pruning on two compositional codebooks.
   \item \textbf{LEGCF} \cite{liang2024lightweight}: A state-of-the-art compositional embedding method that utilizes a codebook and a learnable assignment matrix to infer the full embedding table. \workname is an improved work of LEGCF.
   \item \textbf{Post4bits}: \cite{guan2019post}: A post-training quantization method that devises the greedy search algorithm to find the best min/max bound for quantization scaling.
\end{itemize}

Moreover, we also conduct experiments using the unified dimensionality setting (\textbf{UD)} with dimension sizes of $64$ and $128$ to discuss the effect of \workname \text{} in retaining the recommendation performance amid massive storage and runtime computation complexity reduction.

\subsubsection{Evaluation Metrics}
We use the common recommendation ranking quality metrics \textbf{NDCG@N} \cite{wang2013theoretical} and \textbf{Recall@N} with $N$ set to $\{10, 20\}$ to evaluate the performance of \workname \text{} and baselines mentioned above. We also indicate the \textbf{storage cost} of the resultant embedding tables and the \textbf{peak runtime memory consumption} generated by each work to compare their robustness in fulfilling tight storage constraints.  
\subsubsection{Implementation Details}
In terms of the dimensional size of embedding vectors, for \workname \text{} and all baselines other than Post4bits, we set dimension size $d$ of embeddings/meta-embeddings to $128$. Since Post4bits only supports 4-bit quantization, we first calculate the storage cost of the embedding table generated by our work and then convert it to a suitable dimension size for Post4bits so that the total storage cost is similar to ours for fair comparison. 

For \workname, in default settings, the number of meta-embeddings in the quantized codebook $c$ is set to $2,000$ for Yelp2020 and Amazon-book and $10,000$ for iFashion. We set the quantization bit length $b = 16$, which means each meta-embedding on $\textbf{E}_{meta}$ will be quantized to an INT16 vector. 
By default, we set the retention ratio to $0.7$ (\ie $m = \lfloor 0.7N \rfloor$). We set the maximum neighbor hops for graph rewiring $T = 4$ for all three datasets. We set the predefined boundary for linear integer programming $o = 0.5$. For both pretraining and fine-tuning processes, the number of GNN propagation layers $L$ is set to 4. The number of placeholder meta-embeddings in $\textbf{C}_{prune}$ is set $r = 500$ for Yelp2020 and Amazon-book and $r = 2000$ for the large-scale dataset iFashion. 
The study of the impact of the number of quantized meta-embedding $c$ is presented in Fig. \ref{fig:quant_codebook_num}. The study of the number of the placeholder meta-embedding $r$ is available in Fig. \ref{fig:placeholder_emb_num}.
For the training process, we devise the Adam optimizer \cite{kingma2014adam} with a learning rate $\expnumber{1}{-3}$ and weight decay $\expnumber{1}{-5}$. The $L_2$ penalty factor $\lambda$ is fixed to $\expnumber{5}{-4}$. For Yelp2020 and Amazon-book datasets, we train and test our work on a workstation equipped with an Intel Core i7-12700K processor and an NVIDIA GeForce RTX 3090 GPU. For iFashion dataset, we use an HPC equipped with an AMD EPYC 7443 processor and an NVIDIA GeForce H100 GPU. We set a RAM upper bound of 128GB and a maximum wall time of 7 days for each running process in every work to avoid overly extensive computation. 

To create a fair comparison between our work and baselines, we first calculate the storage cost of the embedding table generated by our work and then input the cost as the storage target for work that accommodates various storage cost requirements. For work that does not offer the option to customize storage targets, like ESAPN \cite{liu_automated_2020} and OptEmbed  \cite{lyu_optembed_2022}, we tune the candidate embedding size set to reach a balanced trade-off between the final storage cost and recommendation performance.

\subsection{Overall Performance (RQ1)} \label{sec:rq1}
\begin{table}[t!]
    \caption{Performance comparison between our work and baselines. ``Storage''  indicates the total storage cost of the \underline{embedding layer} of each method. ``MEM'' means the peak runtime memory consumption measured in each setting. ``UD - dim 128'' and ``UD - dim 64'' are the full embedding table settings with unified dimensions $d=128$ and $d=64$, respectively. ``---'' indicates the results are not available. In each column, we use bold font and underlined text to mark the best and second best results achieved by lightweight embedding methods respectively.}
    \label{tab:overall_performance}
\centering
\resizebox{\textwidth}{!}{
\setlength\tabcolsep{1pt}
\begin{tabular}{c|cccccc|cccccc|cccccc}
\toprule
             & \multicolumn{6}{c|}{Yelp2020}                                                            & \multicolumn{6}{c|}{Amazon-book}                                                         & \multicolumn{6}{c}{iFashion}                                                               \\ \hline
Method       & Storage & MEM    & N@10            & R@10            & N@20            & R@20            & Storage & MEM    & N@10            & R@10            & N@20            & R@20            & Storage  & MEM     & N@10            & R@10            & N@20            & R@20            \\ \hline
UD - dim 128 & 56.74MB & 1.15GB & 0.0284          & 0.0426          & 0.0382          & 0.0721          & 70.43MB & 1.66GB & 0.0172          & 0.0215          & 0.0230          & 0.0367          & 760.46MB & 26.22GB & 0.0053          & 0.0079          & 0.0077          & 0.0145          \\
UD - dim 64  & 28.37MB & 0.99GB & 0.0274          & 0.0409          & 0.0366          & 0.0687          & 35.22MB & 1.28GB & 0.0165          & 0.0204          & 0.0222          & 0.0353          & 380.23MB & 15.81GB & 0.0048          & 0.0070          & 0.0069          & 0.0128          \\ \hline
ESAPN        & 4.29MB  & 0.77GB & 0.0067          & 0.0086          & 0.0087          & 0.0142          & 5.49MB  & 1.06GB & 0.0045          & 0.0037          & 0.0051          & 0.0057          & 37.50MB  &  19.50GB       & 0.0019          & 0.0016          & 0.0023          & 0.0029          \\
OptEmbed     & 1.85MB  & 2.02GB & 0.0084          & 0.0089          & 0.0108          & 0.0158          & 6.55MB  & 2.37GB & 0.0037          & 0.0032          & 0.0050          & 0.0062          & ---      & ---     & ---             & ---             & ---             & ---             \\
PEP          & 1.76MB  & 1.47GB & 0.0185          & 0.0208          & 0.0235          & 0.0351          & 2.01MB  & 1.71GB & 0.0047          & 0.0050          & 0.0058          & 0.0081          & 21.55MB  & 31.87GB & 0.0001          & 0.0001          & 0.0002          & 0.0003          \\
CERP         & 1.76MB  & 1.06GB & 0.0149          & 0.0214          & 0.0203          & 0.0375          & 2.01MB  & 1.48GB & 0.0092          & 0.0112          & 0.0125          & 0.0198          & 17.12MB  & 30.70GB & 0.0012          & 0.0013          & 0.0016          & 0.0025          \\
CIESS        & 1.85MB  & 2.12GB & 0.0178          & {\ul 0.0306}    & 0.0241          & {\ul 0.0521}    & 2.12MB  & 2.75GB & 0.0041          & 0.0046          & 0.0055          & 0.0084          & ---      & ---     & ---             & ---             & ---             & ---             \\
Post4bits    & 1.77MB  & 1.05GB & 0.0191          & 0.0274          & 0.0253          & 0.0463          & 2.06MB  & 1.46GB & 0.0132          & 0.0152          & {\ul 0.0172}    & {\ul 0.0261}    & 17.82MB  & 22.57GB & 0.0014          & 0.0016          & 0.0019          & 0.0028          \\
LEGCF        & 1.76MB  & 1.09GB & {\ul 0.0194}    & 0.0272          & {\ul 0.0260}    & 0.0475          & 2.01MB  & 1.48GB & {\ul 0.0142}    & {\ul 0.0161}    & {\ul 0.0172}    & 0.0244          & 17.12MB  & 30.72GB & {\ul 0.0025}    & {\ul 0.0029}    & {\ul 0.0032}    & {\ul 0.0047}    \\
\workname    & 1.76MB  & 0.57GB & \textbf{0.0216} & \textbf{0.0311} & \textbf{0.0294} & \textbf{0.0550} & 2.01MB  & 0.83GB & \textbf{0.0146} & \textbf{0.0165} & \textbf{0.0188} & \textbf{0.0280} & 17.12MB  & 16.54GB & \textbf{0.0045} & \textbf{0.0050} & \textbf{0.0057} & \textbf{0.0081} \\ \bottomrule
\end{tabular}
}
\end{table}

\begin{figure*}[t!]
    \begin{subfigure}[b]{.7\textwidth}
        \includegraphics[width=1.05\textwidth]{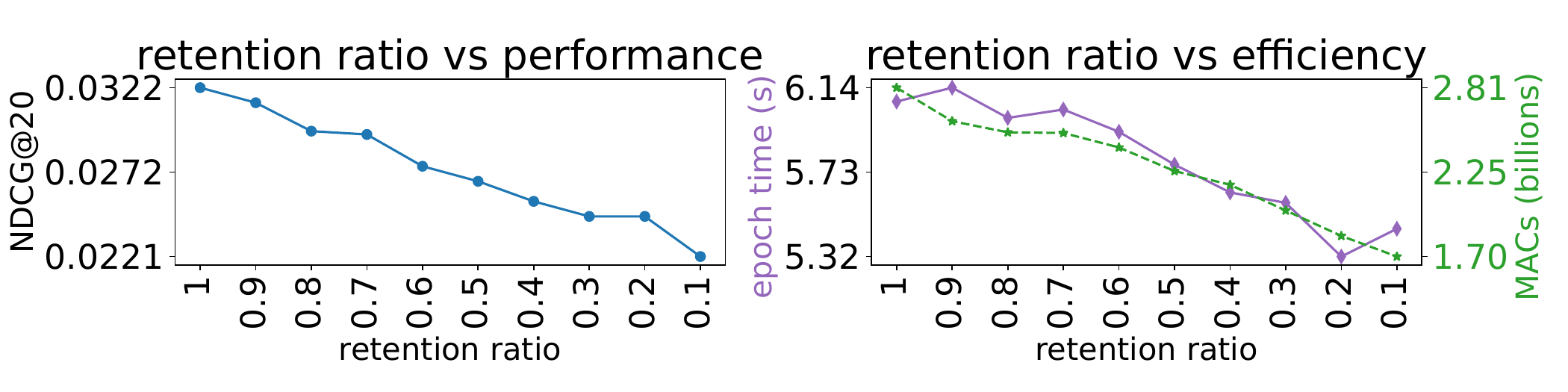}
        \vspace*{-6mm}
        \caption[]%
        {{\small Yelp2020}}
        \label{fig:rq1_yelp}
    \end{subfigure}%
    \vfill
    \begin{subfigure}[b]{.7\textwidth}
        \includegraphics[width=1.05\textwidth]{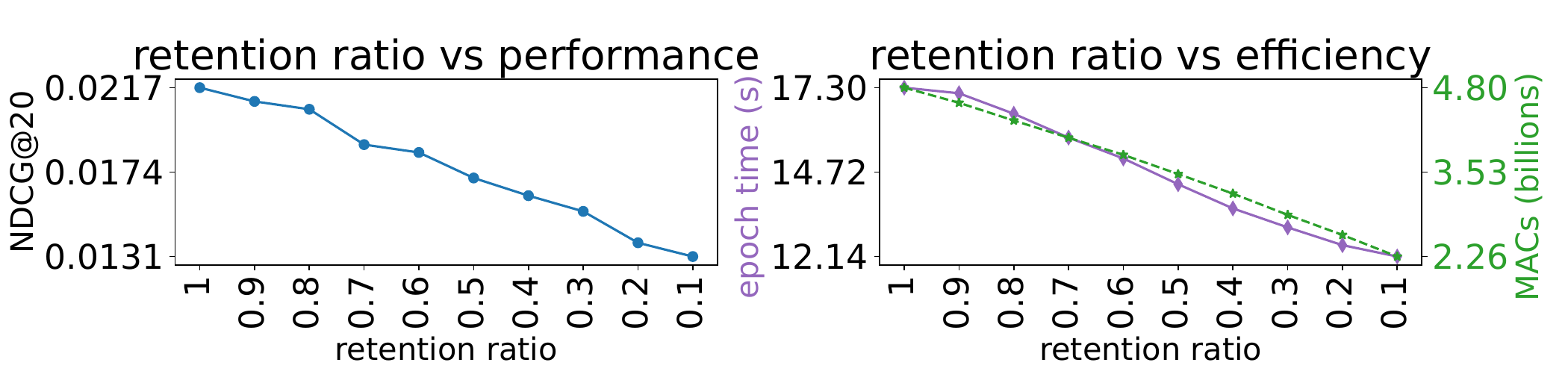}
        \vspace*{-6mm}
        \caption[]%
        {{\small Amazon-book}}
        \label{fig:rq1_amazon}
    \end{subfigure}%
    \vfill
    \begin{subfigure}[b]{.7\textwidth}
        \includegraphics[width=1.05\textwidth]{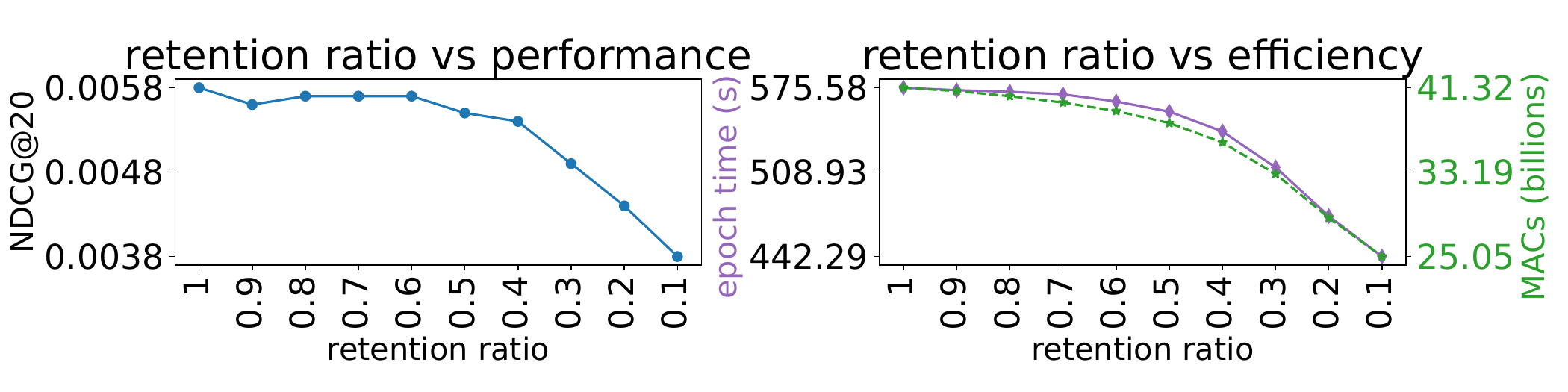}
        \vspace*{-6mm}
        \caption[]%
        {{\small iFashion}}
        \label{fig:rq1_ifashion}
    \end{subfigure}%
    \caption{The plots on the left hand side show the recommendation performance of \workname \text{} w.r.t. different retention ratios. The plots on the right hand side depict the relationship between retention ratios and the computational efficiency of our algorithm. The trend of training epoch elapsed time (in seconds) is shown in purple color. The trend of MACs (in billions) incurred during graph propagation is shown in green color. Here we record the training time per epoch as an indicator of algorithm time efficiency to align with the potential need for fine-tuning the model directly on-edge once deployed.}
    \label{fig:rq1}
\end{figure*}

The overall performance benchmark is revealed in Tab. \ref{tab:overall_performance}. We also plot the relationship between various retention ratios with recommendation performance and graph computational efficiency in Fig. \ref{fig:rq1}. We organize our findings into the following subsections:

\textbf{Comparison with Baselines.} It is observed that \workname \text{} achieves the best performance across all three datasets, followed by our previous work LEGCF, both of which implement a compositional codebook and an assignment weight matrix. This verifies the robustness of compositional codebook structure in enhancing entity embedding uniqueness amid a tight storage budget. Post4Bits and PEP work reasonably well on datasets with fewer users and items such as Yelp2020. On iFashion dataset, however, both work fails to maintain a satisfactory performance. The cause of poor performance may be the overly sparse embedding table/small quantization precision bit length that limits embedding expressiveness. Comparing the results of \workname \text{} with UD settings, our work aces in retaining a robust recommendation performance while reducing the storage cost of the embedding table to a tiny proportion of full embedding tables. This observation is especially obvious on the largest dataset iFashion, as the storage cost of \workname \text{} is reduced by $97.7\%$ of that consumed by the UD setting with a dimension size of 128, yet the performance is only degraded by $15.1\%$ when evaluated using NDCG@10. In terms of the storage cost of baselines, AutoML-based dimension search methods like ESAPN and OptEmbed tend to obtain a much larger embedding table. The possible cause of this is that these methods assign higher priority to optimizing the performance over memory cost. In terms of the peak runtime memory consumption, our work exerts the lowest volume of memory across all three datasets. Interestingly, it is discovered that some embedding optimization work has an even higher peak runtime memory consumption than the UD settings, creating excessive burdens in the pipeline. We also notice that AutoML-based methods OptEmbed and CIESS fail to generate recommendation results on the largest dataset iFashion, due to long elapsed time and/or excessive runtime memory consumption. This phenomenon proves the statement that the AutoML-based methods are not suitable for large-scale deployment.

\textbf{Retention Ratio vs Performance/Efficiency.} To measure the training epoch time, we fix the training batch for each dataset for settings with various retention ratios. To measure the MACs incurred during the graph propagation operation, we first generate propagation graphs with various ratios of entities retained, and then conduct the calculation of multiply-addictive counts on each propagation graph. It is observed that for smaller datasets Yelp2020 and Amazon-book, the recommendation performance is relatively sensitive w.r.t. the retention ratio. On iFashion, \workname \text{} still manages to maintain a satisfactory performance when $40\%$ of the total entities are retained, creating an optimal trade-off between recommendation performance and graph computational efficiency. In terms of the relationship between retention ratios and algorithm efficiency, it is noticed that as more entities are pruned, the sparsity of the rewired propagation graph increases, followed by fewer computation costs (\ie lower MACs). The epoch training time also drops w.r.t. to retention ratios, due to lighter graph computational overhead and fewer entity embeddings to undergo backward propagation. The experiments conducted in this subsection verify \workname's ability to fit state-of-the-art GNN-based recommender systems to resource-constrained edge devices.

\subsection{Discussions on Key Model Components (RQ2)}
\begin{table*}[t!]
    \caption{Performance comparison between default settings and settings with a particular component altered. The default settings are the ones fully trained on the quantized compositional codebook. The BIP constraint is solved to identify contributions of entities in graph propagation for graph sparsification. In the fine-tuning stage, graph-propagated embeddings of pruned entities are drawn from the placeholder meta-embeddings $\textbf{C}_{prune}$ with $\textbf{H}_{prune}$ discarded.}
    \label{tab:ablation}
\centering
\resizebox{\textwidth}{!}{
\setlength\tabcolsep{4pt}
\begin{tabular}{cl|cccc|cccc|cccc}
\toprule
          \multicolumn{2}{c|}{Dataset}                                                              & \multicolumn{4}{c|}{Yelp2020}     & \multicolumn{4}{c|}{Amazon-book}  & \multicolumn{4}{c}{iFashion}      \\ \hline
\multicolumn{1}{c|}{Retention Ratio}      & Variant                       & N@10   & R@10   & N@20   & R@20   & N@10   & R@10   & N@20   & R@20   & N@10   & R@10   & N@20   & R@20   \\ \hline
\multicolumn{1}{c|}{\multirow{6}{*}{0.7}} & Default                       & 0.0216 & 0.0311 & 0.0294 & 0.0550 & 0.0146 & 0.0165 & 0.0188 & 0.0280 & 0.0045 & 0.0050 & 0.0057 & 0.0081 \\
\multicolumn{1}{c|}{}                     & w/o Fine-tuning               & 0.0214 & 0.0309 & 0.0291 & 0.0545 & 0.0139 & 0.0161 & 0.0182 & 0.0278 & 0.0043 & 0.0049 & 0.0055 & 0.0079 \\
\multicolumn{1}{c|}{}                     & w/o BIP                       & 0.0218 & 0.0309 & 0.0287 & 0.0521 & 0.0129 & 0.0147 & 0.0165 & 0.0245 & 0.0038 & 0.0041 & 0.0048 & 0.0065 \\
\multicolumn{1}{c|}{}                     & w/o $\textbf{C}_{prune}$      & 0.0164 & 0.0235 & 0.0222 & 0.0415 & 0.0116 & 0.0126 & 0.0142 & 0.0202 & 0.0046 & 0.0049 & 0.0057 & 0.0079 \\
\multicolumn{1}{c|}{}                     & $\textbf{H}_{prune}$ retained & 0.0166 & 0.0228 & 0.0247 & 0.0440 & 0.0142 & 0.0180 & 0.0159 & 0.0263 & 0.0044 & 0.0055 & 0.0049 & 0.0080 \\
\multicolumn{1}{c|}{}                     & Post-compression              & 0.0136 & 0.0207 & 0.0190 & 0.0374 & 0.0101 & 0.0116 & 0.0132 & 0.0202 & 0.0029 & 0.0032 & 0.0038 & 0.0055 \\ \hline
\multicolumn{1}{c|}{\multirow{6}{*}{0.5}} & Default                       & 0.0190 & 0.0279 & 0.0266 & 0.0514 & 0.0132 & 0.0151 & 0.0171 & 0.0258 & 0.0044 & 0.0048 & 0.0055 & 0.0078 \\
\multicolumn{1}{c|}{}                     & w/o Fine-tuning               & 0.0182 & 0.0269 & 0.0256 & 0.0495 & 0.0124 & 0.0146 & 0.0164 & 0.0254 & 0.0042 & 0.0047 & 0.0054 & 0.0076 \\
\multicolumn{1}{c|}{}                     & w/o BIP                       & 0.0194 & 0.0269 & 0.0256 & 0.0457 & 0.0131 & 0.0148 & 0.0167 & 0.0247 & 0.0033 & 0.0034 & 0.0041 & 0.0055 \\
\multicolumn{1}{c|}{}                     & w/o $\textbf{C}_{prune}$      & 0.0121 & 0.0170 & 0.0166 & 0.0310 & 0.0095 & 0.0099 & 0.0116 & 0.0160 & 0.0043 & 0.0045 & 0.0054 & 0.0073 \\
\multicolumn{1}{c|}{}                     & $\textbf{H}_{prune}$ retained & 0.0146 & 0.0200 & 0.0215 & 0.0381 & 0.0139 & 0.0174 & 0.0156 & 0.0255 & 0.0044 & 0.0056 & 0.0049 & 0.0080 \\
\multicolumn{1}{c|}{}                     & Post-compression              & 0.0118 & 0.0175 & 0.0165 & 0.0320 & 0.0099 & 0.0113 & 0.0130 & 0.0198 & 0.0029 & 0.0032 & 0.0038 & 0.0055 \\ \hline
\multicolumn{1}{c|}{\multirow{6}{*}{0.1}} & Default                       & 0.0157 & 0.0232 & 0.0221 & 0.0427 & 0.0097 & 0.0108 & 0.0131 & 0.0199 & 0.0031 & 0.0032 & 0.0038 & 0.0052 \\
\multicolumn{1}{c|}{}                     & w/o Fine-tuning               & 0.0129 & 0.0194 & 0.0185 & 0.0365 & 0.0088 & 0.0102 & 0.0120 & 0.0186 & 0.0031 & 0.0032 & 0.0038 & 0.0052 \\
\multicolumn{1}{c|}{}                     & w/o BIP                       & 0.0118 & 0.0175 & 0.0168 & 0.0326 & 0.0072 & 0.0073 & 0.0095 & 0.0135 & 0.0015 & 0.0015 & 0.0018 & 0.0024 \\
\multicolumn{1}{c|}{}                     & w/o $\textbf{C}_{prune}$      & 0.0063 & 0.0085 & 0.0092 & 0.0173 & 0.0041 & 0.0037 & 0.0051 & 0.0066 & 0.0021 & 0.0020 & 0.0025 & 0.0033 \\
\multicolumn{1}{c|}{}                     & $\textbf{H}_{prune}$ retained & 0.0074 & 0.0107 & 0.0113 & 0.0215 & 0.0108 & 0.0138 & 0.0121 & 0.0205 & 0.0036 & 0.0035 & 0.0042 & 0.0052 \\
\multicolumn{1}{c|}{}                     & Post-compression              & 0.0062 & 0.0093 & 0.0093 & 0.0187 & 0.0080 & 0.0091 & 0.0105 & 0.0160 & 0.0026 & 0.0028 & 0.0034 & 0.0048 \\ \bottomrule
\end{tabular}
}
\end{table*}

To validate the effectiveness of key components of \workname, we conduct ablation studies on the embedding fine-tuning process, the BIP procedure of identifying pruned entities for the graph rewiring process, the placeholder meta-embeddings $\textbf{C}_{prune}$ from clustering for pruned entities in the fine-tuning stage, the revocation of $\textbf{H}_{prune}$ for pruned entities in the fine-tuning stage, and the end-to-end training on the quantized compositional codebook, respectively. The experiment results are recorded in Tab. \ref{tab:ablation}. In order to thoroughly study the impact of key components of \workname \text{} w.r.t. different edge device specifications, we display the results conducted with retention ratios set to $\{0.7, 0.5, 0,1\}$ respectively.

\textbf{Embedding Fine-tuning Process.} We denote the settings where the pretrained quantized compositional embedding table is used directly on edge devices without fine-tuning as \textit{w/o Fine-tuning} in Tab. \ref{tab:ablation}. It is obvious that applying the rewired graph directly for inference without the embedding finefining step tends to achieve sub-optimal recommendation performance. The effect of fine-tuning is particularly obvious when evaluated on Yelp2020 dataset with the retained ratio set to $0.1$, as performance discrepancy between the setting with and without fine-tuning shows a $17.8\%$ difference when observing the NDCG@10 values. The experiment results indicate the necessity of a lightweight embedding fintuning process to be conducted on edge devices. 

\textbf{BIP for Graph Rewiring.} We denote the scenarios where the pruned entities are randomly drawn from the entity set as \textit{w/o BIP} in Tab. \ref{tab:ablation}. It is discovered that as the retention ratio decreases, the performance discrepancy between default settings and settings without the use of BIP enlarges on all three datasets. This indicates the importance of solving the BIP constraint for the graph rewiring process as it effectively identifies the entities whose contribution to graph propagation is more impactful than others. In the case that only a small percentage of entities are retained in the rewired graph, the rewired graph maximally preserves the collaborative semantics by retaining entities that are deemed to have significant contributions to graph propagation.

\textbf{$\textbf{C}_{prune}$ for Entity Imputation.} We denote the settings where the pruned entity embeddings in the fine-tuning stage are imputed as the mean of all pruned entity graph-propagated pretrained embeddings as \textit{w/o $\textbf{C}_{prune}$} in Tab. \ref{tab:ablation}. It is observed that on smaller datasets Yelp2020 and Amazon-book, the use of $\textbf{C}_{prune}$ for pruned entity embedding imputation plays a critical role in attaining excellent accuracy. This is especially true when the retention ratio is set to $0.1$. On large-scale dataset iFashion, when the retention ratios are $0.7$ and $0.5$, the performance gap between settings with and without $\textbf{C}_{prune}$ is not particularly noticeable. However, when the retention ratio is lowered to $0.1$, the performance of the setting with $\textbf{C}_{prune}$ for pruned entity embedding imputation prevails. We deduce the reason why using $\textbf{C}_{prune}$ in low retention ratio cases works well is that when the majority of entities are pruned, imputing their embeddings from the placeholder codebook $\textbf{C}_{prune}$ provides sufficient embedding uniqueness for the recommender system to distinguish entities, rather than simply taking the unanimous mean placeholder embedding for all pruned entities.

\textbf{Revocation of $\textbf{H}_{prune}$ in Fine-tuning.} In Sec. \ref{sec:finetuning}, we conclude that the primary reason for deriving the embeddings of pruned entities ($\hat{\textbf{H}}_{prune}$) from the fixed placeholder meta-embeddings $\textbf{C}_{prune}$, rather than updating them like the retained entities, is to significantly reduce runtime memory consumption and computational complexity. To study the impact on recommendation accuracy after discarding pruned entities' full graph-propagated embeddings $\textbf{H}_{prune}$ in favor of $\textbf{C}_{prune}$, we show the results computed in scenarios where $\textbf{H}_{prune}$ were also allowed to participate in the fine-tuning process (denoted as \textit{$\textbf{H}_{prune}$ retained}) in Tab. \ref{tab:ablation}. The indicated results show that this modification leads to only marginal improvements, and in some cases, slightly degraded performance, particularly when the retention ratio is relatively high. The experiments conducted in this sub-section confirm that our design will not significantly affect the performance caused by the different updating methods of pruned and retained entities.

\textbf{End-to-end Training on Quantized Compositional Codebook.} In our work, we initialize the compositional embedding framework and perform quantization-aware training directly on it, as opposed to the work \cite{guan2019post} that performs a post-training quantization process on the well-trained uncompressed full embedding table. To examine the performance superiority of our end-to-end training design, we conduct an additional experiment adopting a post-compression strategy. In this setting, we first train a full embedding table to convergence, and then apply our proposed graph rewiring, meta-embedding generation, and quantization modules as post hoc compression steps. This setup mirrors traditional compression approaches that operate after full-capacity training. The results, reported in Tab. \ref{tab:ablation} under the \textit{Post-compression} setting, indicate a notable degradation in performance compared to our end-to-end training scheme. We hypothesize that the post-compression strategy fails to establish a high-quality and semantically aligned mapping between entities and meta-embeddings, leading to suboptimal compositions. In contrast, our joint learning framework enables better adaptation of the codebook and assignments throughout training.

\subsection{Hyperparameter Sensitivity (RQ3)} \label{sec_rq3}
\begin{figure}[t!] 
    \centering
    \begin{subfigure}[b]{.5\textwidth}
        \includegraphics[width=\textwidth]{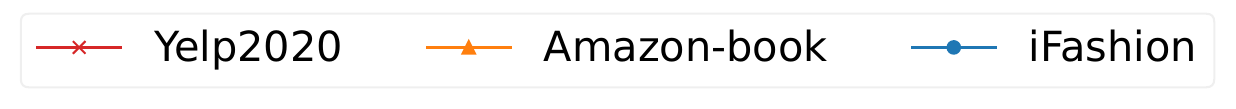}
    \end{subfigure}%
    \vfill
     \begin{subfigure}[b]{.333\textwidth}
        \includegraphics[width=.7\textwidth]{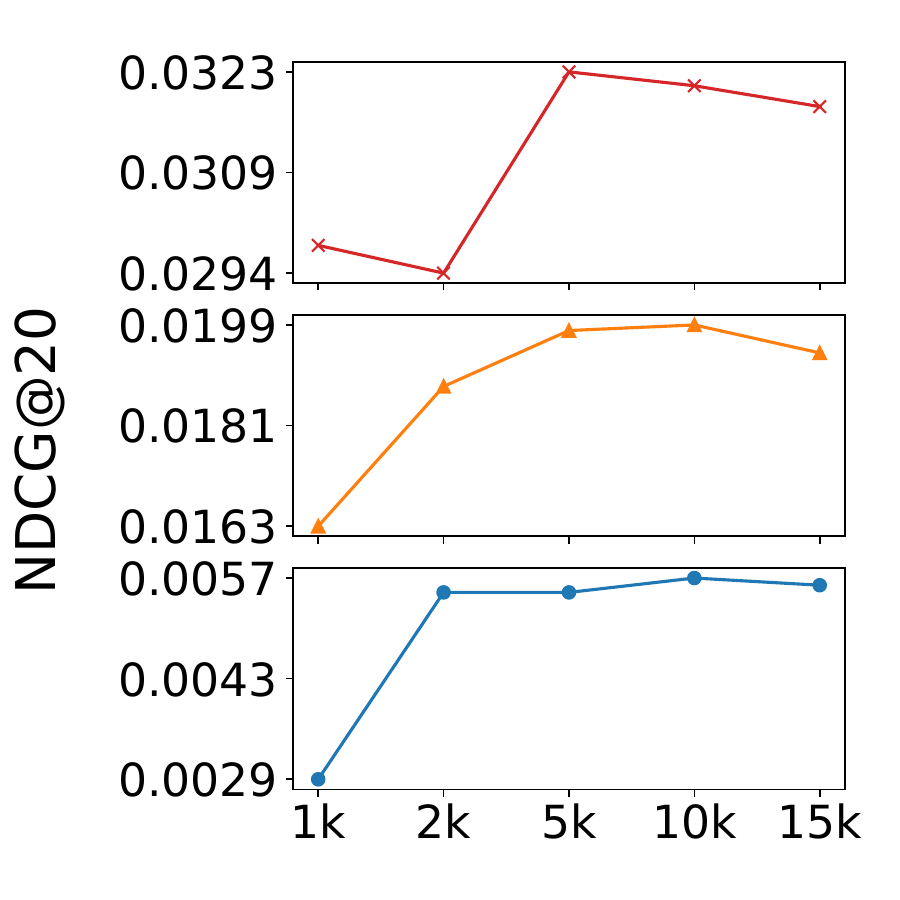}
        \vspace*{-3mm}
        \caption[]%
        {{\small Effect of $c$}}
        \label{fig:quant_codebook_num}
    \end{subfigure}%
    \begin{subfigure}[b]{.333\textwidth}
        \includegraphics[width=.7\textwidth]{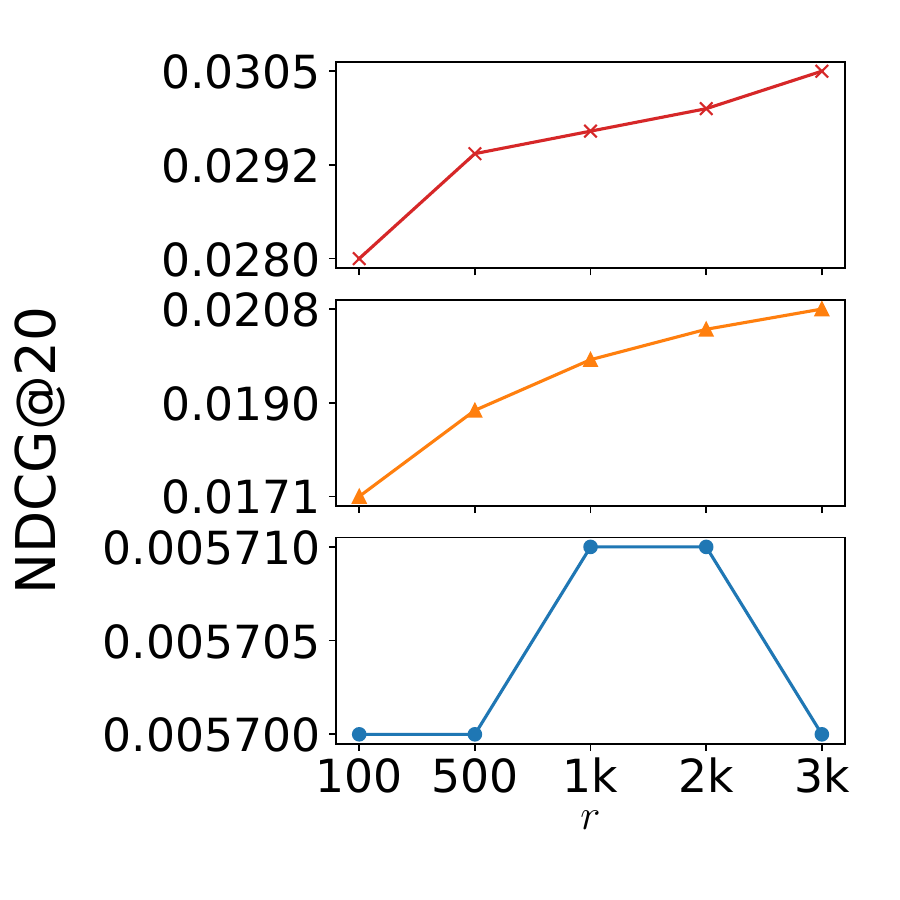}
        \vspace*{-3mm}
        \caption[]%
        {{\small Effect of $r$}}
        \label{fig:placeholder_emb_num}
    \end{subfigure}%
    \begin{subfigure}[b]{.333\textwidth}
        \includegraphics[width=.7\textwidth]{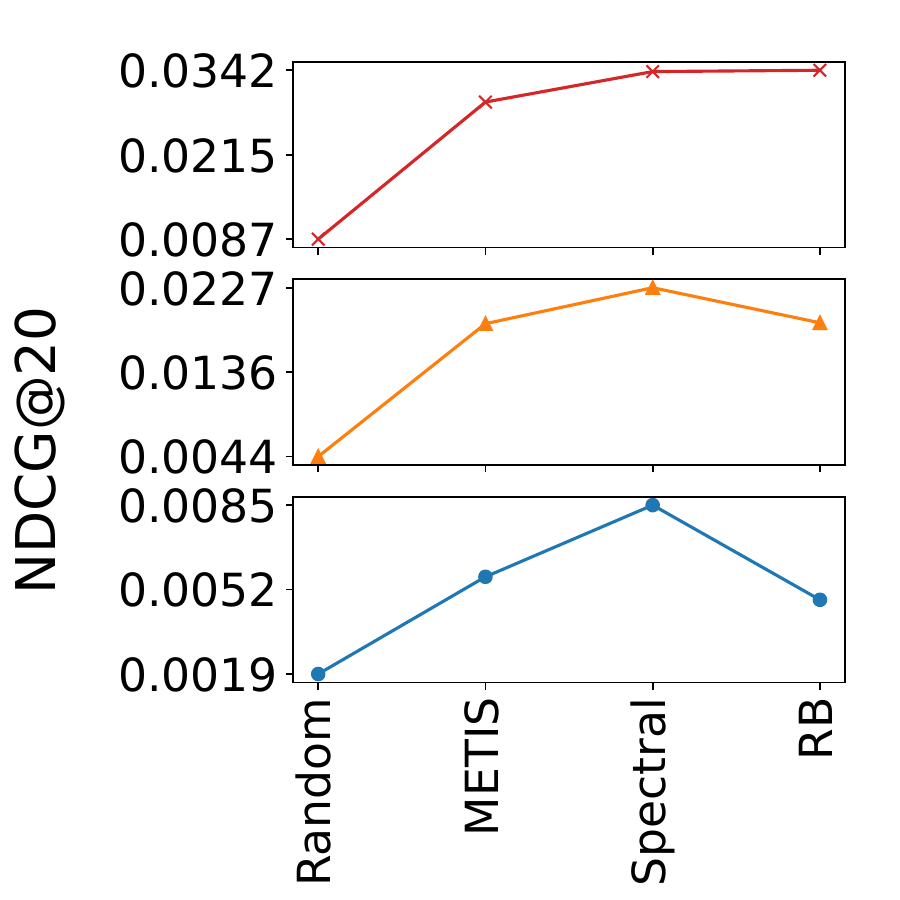}
        \vspace*{-3mm}
        \caption[]%
        {{\small Effect of assignment init. method}}
        \label{fig:assignment_init}
    \end{subfigure}%
    \caption{The performance of \workname \text{} w.r.t. various hyperparameter settings.}
    \label{fig:hyperparams}
\end{figure}

The three tunable hyperparameters in \workname \text{} are the number of meta-embeddings in the quantized codebook $c$, the number of placeholder meta-embeddings for pruned entity embedding imputation $r$, as well as the initialization method of the assignment matrix.

\textbf{Number of Quantized Meta-Embeddings $c$.} The impact of varying number of meta-embeddings $c$ in the quantized codebook $\Bar{\textbf{E}}_{meta}$ is revealed in Fig. \ref{fig:quant_codebook_num}. The set of testing value for $c$ include $\{1,000; 2,000; 5,000; 10,000; 15,000\}$. The performance plot shows that setting $c = 5,000$ yields optimal performance on all three datasets. Setting $c$ to values larger than $5,000$ does not lead to great performance improvement. We also notice that for the large-scale dataset iFashion, setting $c = 2,000$ results in even better performance than setting $c = 5,000$. Based on the two observations, we conclude that \workname \text{} is designed to work under a tight storage cost constraint for large-scale datasets, effectively easing the deployment difficulty of GNN-based recommender systems on resource-constrained devices while maintaining satisfactory recommendation performance. 

\textbf{Number of Placeholder Meta-Embeddings $r$.} The effect of different placeholder meta-embeddings $r$ in the codebook $\textbf{C}_{prune}$ is visualized in Fig. \ref{fig:placeholder_emb_num}. We examine the settings with $r$ set to $\{100; 500; 1,000; 2,000; 3,000\}$. It is discovered that for smaller datasets Yelp2020 and Amazon-book, increasing $r$ yields better recommendation performance. However, for large-scale dataset iFashion, changing the value of $r$ barely causes fluctuations in the performance trend. The cause of this might be that iFashion contains a large number of entities that actually have very few interactions with other entities. As such, in settings with a relatively high retention ratio (\ie $0.7$), the size of the pruned entities is too small such that the placeholder embeddings are barely drawn for downstream recommendation tasks. Hence, the little accuracy change is spotted despite using various sizes of placeholder meta-embeddings. Our experiments in this section indicate that the number of placeholder meta-embeddings for pruned entities is not a key performance factor for large-scale datasets. One can simply opt for a small $r$ value for large-scale datasets so the storage cost of the placeholder meta-embeddings is negligible. 

\textbf{Initialization Method of Assignment Matrix.} The effect of various methods used to select the anchor meta-embedding on initialization is shown in Fig. \ref{fig:assignment_init}. We showcase the performance results conducted under 4 initialization strategies: random selection (dubbed ``Random''), METIS k-way graph partitioning (dubbed ``METIS'') \cite{karypis1997metis}, spectral clustering (dubbed ``Spectral'') \cite{von2007tutorial} and METIS recursive bisection graph partitioning (dubbed ``RB'') \cite{chan1993spectral}. It is obvious that while the performance achieved by the three graph partitioning algorithms varies, the magnitude of change is barely noticeable across the three initialization methods. In contrast, the performance achieved by randomly selecting an anchor meta-embedding is much worse than the settings with a graph partitioning algorithm in use. This implies that the main factor of a robust meta-embedding-based recommender system is the proper initialization of the assignment matrix. With a reasonably well anchor meta-embedding assignment, our framework is capable of obtaining satisfactory accuracy without further refinement of the assignment matrix.

\subsection{Framework Generalizability (RQ4)}\label{sec_rq4}
\begin{table*}[t!]
    \caption{Performance comparison between our work and unified dimensionality settings. ``\textit{recommender} - \workname \text{} - \textit{1}/\textit{0.7}" means the setting evaluated on our work with all/$70\%$ of the entities retained when the \textit{recommender} is used. ``\textit{recommender} - UD'' means the setting evaluated on unified dimensionality when the \textit{recommender} is used.}
    \label{tab:graph_generalizability}
\centering
\setlength\tabcolsep{3pt}
\resizebox{\textwidth}{!}{
\begin{tabular}{l|cccc|cccc|cccc}
\toprule
                           & \multicolumn{4}{c|}{Yelp2020}     & \multicolumn{4}{c|}{Amazon-book}  & \multicolumn{4}{c}{iFashion}      \\ \hline
Setting                    & N@10   & R@10   & N@20   & R@20   & N@10   & R@10   & N@20   & R@20   & N@10   & R@10   & N@20   & R@20   \\ \hline
LightGCN - \workname \text{} - 1   & 0.0241 & 0.0347 & 0.0322 & 0.0596 & 0.0168 & 0.0195 & 0.0217 & 0.0330 & 0.0046 & 0.0051 & 0.0058 & 0.0082 \\
LightGCN - \workname \text{} - 0.7 & 0.0216 & 0.0311 & 0.0294 & 0.0550 & 0.0146 & 0.0165 & 0.0188 & 0.0280 & 0.0045 & 0.0050 & 0.0057 & 0.0081 \\
LightGCN - UD              & 0.0157 & 0.0226 & 0.0208 & 0.0386 & 0.0080 & 0.0092 & 0.0106 & 0.0164 & 0.0014 & 0.0016 & 0.0019 & 0.0028 \\ \hline
NGCF - \workname \text{} - 1       & 0.0088 & 0.0121 & 0.0136 & 0.0264 & 0.0107 & 0.0121 & 0.0132 & 0.0190 & 0.0039 & 0.0038 & 0.0042 & 0.0049 \\
NGCF - \workname \text{} - 0.7     & 0.0086 & 0.0119 & 0.0131 & 0.0253 & 0.0083 & 0.0094 & 0.0101 & 0.0146 &  0.0011      &   0.0013     &   0.0015     &   0.0022     \\
NGCF - UD                  & 0.0046 & 0.0071 & 0.0068 & 0.0137 & 0.0051 & 0.0056 & 0.0067 & 0.0101 & 0.0000 & 0.0000 & 0.0001 & 0.0001 \\ \hline
xSimGCL - \workname \text{} - 1    & 0.0232 & 0.0325 & 0.0306 & 0.0550 & 0.0213 & 0.0246 & 0.0275 & 0.0416 & 0.0084 & 0.0088 & 0.0104 & 0.0142 \\
xSimGCL - \workname \text{} - 0.7  & 0.0173 & 0.0247 & 0.0235 & 0.0434 & 0.0170 & 0.0198 & 0.0222 & 0.0338 & 0.0052 & 0.0056 & 0.0066 & 0.0092 \\
xSimGCL - UD               & 0.0163 & 0.0230 & 0.0216 & 0.0395 & 0.0088 & 0.0100 & 0.0115 & 0.0174 & 0.0022 & 0.0026 & 0.0030 & 0.0045 \\ \bottomrule
\end{tabular}}
\end{table*}

Since \workname \text{} is designed to enable the deployment of lightweight GNN-based recommender systems on resource-constrained edge devices, it is necessary to check whether the proposed framework performs well on various GNN-based recommenders. To do this, we select three popular GNN-based recommenders, namely LightGCN \cite{he_lightgcn_2020}, NGCF \cite{wang_neural_2019} and the self-supervised contrastive learning recommender xSimGCL \cite{yu2023xsimgcl}, to carry out performance benchmark in this section. For each base recommender, we reveal the performance achieved by our work when all entities are retained, $70\%$ of the entities are retained, as well as the settings evaluated on the full embedding table (UD) with dimension sizes adjusted to meet similar storage cost to our settings. The results are revealed in Tab. \ref{tab:graph_generalizability}.

From the table, it is noticed that recommendation accuracy of \workname \text{} consistently outperforms the UD settings across all three datasets, regardless of the choice of base recommenders. One extreme case is when NGCF is adopted as the base recommender on iFashion, the accuracy attained by the UD setting is ill-defined. In contrast, the setting using our work with $70\%$ of entities retained still maintains a satisfactory recommendation quality. This indicates that the quantized compositional embedding table, along with the graph rewiring technique, is tailored for GNN-based recommender systems. Regarding the performance drop caused by graph rewiring, it is discovered that when \workname \text{} is equipped with LightGCN or NGCF base recommenders, the performance degradation is not obvious on Yelp2020 and iFashion datasets. The settings with xSimGCL base recommender seem to be more sensitive to retention ratio, although the performance achieved by settings with retention ratio of $0.7$ is still much better than the UD settings. Our guess is that xSimGCL itself implements a graph perturbation technique. Using the rewired propagation graph implements another form of graph perturbation. The two techniques together may have caused slight interference, which leads to sensitivity to retention ratios.

\subsection{Impact of Quantization Precision (RQ5)}
\begin{figure*}[t!]
    \centering
    \begin{subfigure}[b]{.3\textwidth}
        \centering
        \includegraphics[width=\textwidth]{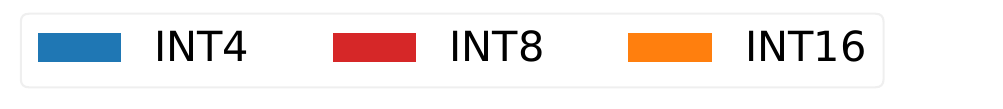}
    \end{subfigure}%
    \vfill
    \begin{subfigure}[b]{.333\textwidth}
        \centering
        \includegraphics[width=.8\textwidth]{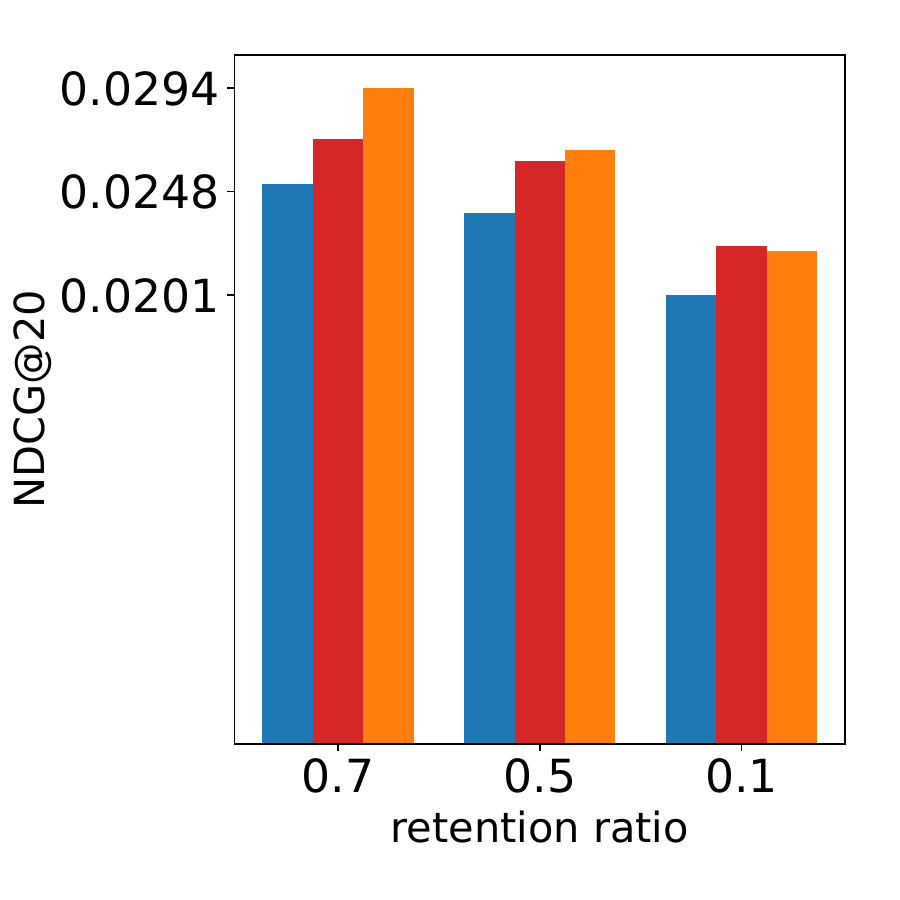}
        \vspace*{-3mm}
        \caption[]%
        {{\small Yelp2020}}
        \label{fig:quant_precision_yelp}
    \end{subfigure}%
    \begin{subfigure}[b]{.333\textwidth}
        \centering
        \includegraphics[width=.8\textwidth]{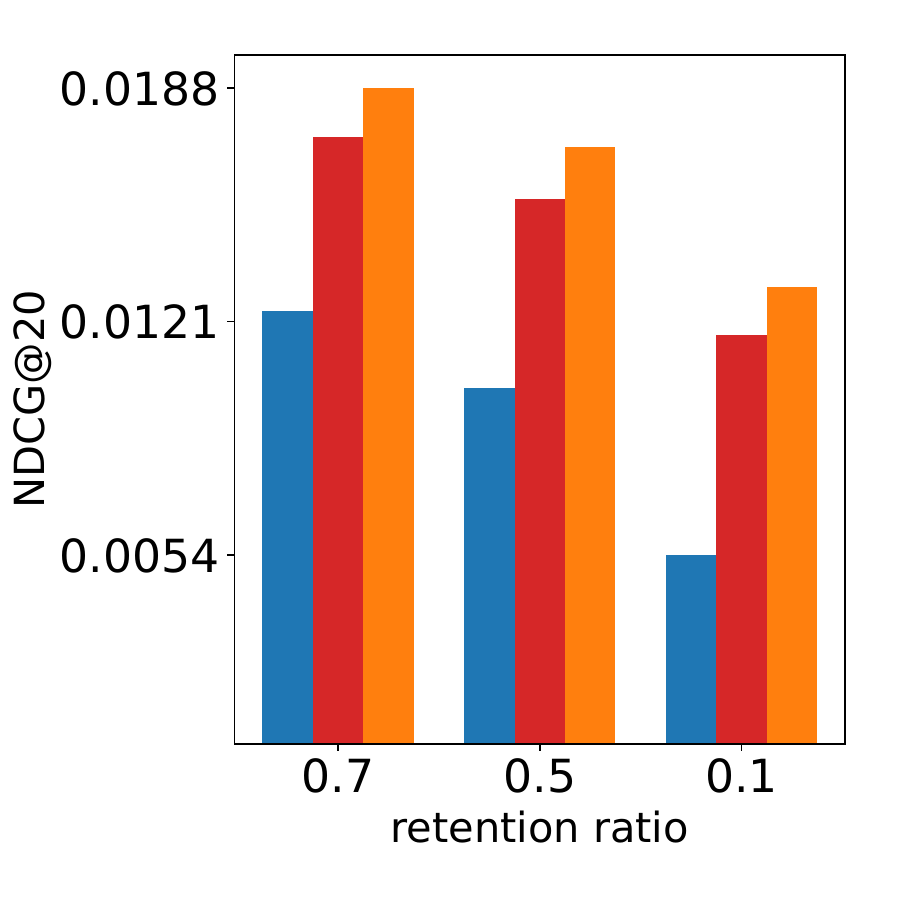}
        \vspace*{-3mm}
        \caption[]%
        {{\small Amazon-book}}
        \label{fig:quant_precision_amazon}
    \end{subfigure}%
    \begin{subfigure}[b]{.333\textwidth}
        \centering
        \includegraphics[width=.8\textwidth]{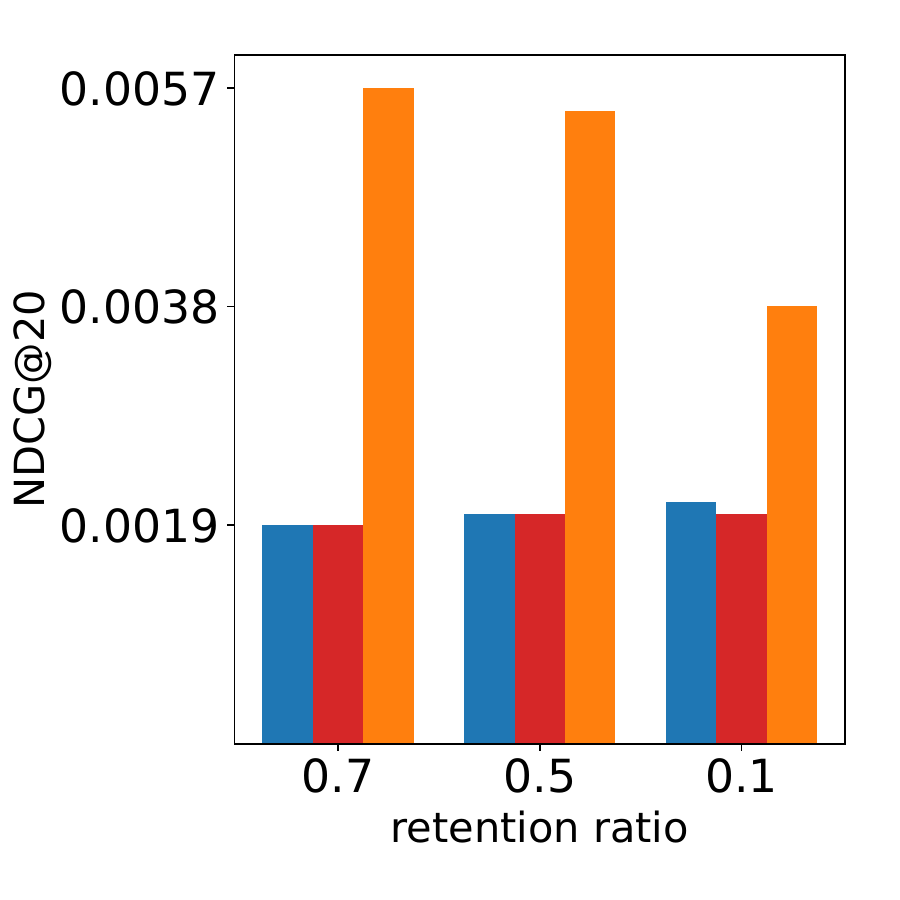}
        \vspace*{-3mm}
        \caption[]%
        {{\small iFashion}}
        \label{fig:quant_precision_ifashion}
    \end{subfigure}%
    \caption{The performance of \workname \text{} when using different quantization precisions for $\Bar{\textbf{E}}_{meta}$.}
    \label{fig:quant_precision}
\end{figure*}

In \workname, we deploy a quantized compositional codebook $\Bar{\textbf{E}}_{meta}$ to further reduce the storage cost of the embedding table. Commonly, the elements in the codebook are quantized into INT8 or INT16 precision due to the native hardware support by most edge devices. Recent quantization literature has pushed down the quantization precision to 4 bits \cite{wu2023understanding, xi2023training, sun2020ultra}, despite most of them defining a self-implemented format and GEMM for storage and computation. From our perspective, it is insightful to study the relationship between the recommendation performance of our work and the choice of quantization precisions applied to the compositional codebook. To do this, we conduct experiments using the quantized compositional codebook $\Bar{\textbf{E}}_{meta}$ with precision formats INT4, INT8 and INT16 respectively. We once again show the results evaluated on retention ratios of $\{0.7, 0.5, 0.1\}$ for a comprehensive study of the subject. The performance results can be found in Fig. \ref{fig:quant_precision}.

From the plots, it is inspected that as the retention ratio decreases, the recommendation performance, in general, follows the same fashion, regardless of the quantization precision formats. It is also discovered that INT4 settings are more sensitive to retention ratios on Yelp2020 and Amazon-book datasets. On the other hand, INT8 and INT16 settings attain similar performance on these two datasets, indicating using INT8 for meta-embedding compression should suffice for smaller datasets. However, on iFashion dataset, the performance gap between INT4 and INT8 settings is tiny. The recommendation performance achieved by INT16 settings shows a great leap forward compared to INT4 and INT8 settings. This means that for large-scale datasets, using a small precision format may limit the expressiveness of the meta-embeddings, and so do entity embeddings.

\subsection{On-device Fine-tuning Computation Improvement (RQ6)}

\begin{table*}[t!]
\caption{Batch-wise runtime efficiency comparison between the pretraining process and settings with different retention ratios. The settings with a retention ratio of 1 correspond to the pretraining process. ``MEM'' is the total runtime memory consumption. ``MFLOPS'' is the magnitude of the exerted megaFLOPS. ``Time'' is the training elapsed time. The three computation-related metrics are measured for a forward and backward pass that occurred in a single batch.}
\label{tab:computation_cost}
\centering
\resizebox{\textwidth}{!}{
\setlength\tabcolsep{4pt}
\begin{tabular}{c|ccc|ccc|ccc}
\toprule
                & \multicolumn{3}{c|}{Yelp2020} & \multicolumn{3}{c|}{Amazon-book} & \multicolumn{3}{c}{iFashion} \\ \hline
Retention Ratio & MEM      & MFLOPS  & Time     & MEM       & MFLOPS   & Time      & MEM     & MFLOPS  & Time     \\ \hline
1               & 52.45MB  & 58.54   & 20.81ms  & 74.06MB   & 77.39    & 37.05ms   & 0.80GB  & 944.82  & 513.39ms \\
0.7             & 48.56MB  & 55.05   & 19.78ms  & 66.40MB   & 63.12    & 33.20ms   & 0.79GB  & 939.22  & 510.74ms \\
0.5             & 37.83MB  & 38.71   & 18.75ms  & 61.62MB   & 60.99    & 28.77ms   & 0.63GB  & 767.26  & 495.65ms \\
0.1             & 33.84MB  & 36.57   & 16.40ms  & 45.99MB   & 56.62    & 22.14ms   & 0.58GB  & 767.26  & 374.26ms \\ \bottomrule
\end{tabular}
}
\end{table*}

Since \workname \text{} offers the option to perform a lightweight embedding fine-tuning process on-edge, it is crucial to study the runtime computation cost of \workname in the fine-tuning stage. To do this, we simulate the pretraining and fine-tuning processes using a pure GPU computation environment. For the three datasets, we fix the batch size and record the runtime memory usage, the measured floating point operations per second (FLOPS) and the elapsed time to perform a forward and backward pass in a single batch. Note that since the main innovation of our work lies in the improvement of the embedding update strategy, the three computation efficiency metrics are measured on the graph propagation operation in the forward and backward passes only. To be specific, we track the accumulative computation cost incurred during the low-level multiplication and accumulation operations. The result comparison is shown in Tab. \ref{tab:computation_cost}. It is evident that, for the three datasets, the pertaining process always incurs the most expensive computation cost and the longest training elapsed time. For fine-tuning, as the retention ratio is lowered, the corresponding runtime memory consumption and training elapsed time is also reduced. It is also apparent that our work works well for large-scale datasets in optimizing the runtime computation cost. This is shown in the iFashion dataset, as the runtime memory consumption and training elapsed time of the setting are respectively $27.1\%$ and $27.7\%$ lower than that measured on the pertaining process with a retention ratio of 0.1. Thus, compared with pretraining, the fine-tuning process is more computationally efficient, thus practically supporting on-device deployment.

\section{Conclusion} \label{sec:conclusion}
In this paper, we analyze the deployment of GNN-based recommender systems on resource-constrained edge devices and conclude the inherited challenges of high embedding storage cost and low graph computational efficiency. We also study currently available embedding optimization frameworks and realize that most work fails to address graph computational efficiency and high peak runtime memory consumption issues. A storage- and memory-efficient GNN-based framework \workname \text{} is then proposed by means of alleviating the difficulty of employing state-of-the-art GNN-based recommender systems on resource-constrained edge devices. We conduct extensive experiments to validate the effectiveness of \workname \text{} in lowering the embedding storage cost and graph computational complexity while preserving satisfactory recommendation performance.

\section*{Acknowledgment}
The Australian Research Council partially supports this work under the streams of Future Fellowship (Grant No. FT210100624), Discovery Early Career Researcher Award (Grant No. DE230101033), Discovery Project (Grant No. DP240101108 and DP240101814), and Linkage Project (Grant No. LP230200892 and LP240200546).


\bibliographystyle{ACM-Reference-Format}
\bibliography{customized}


\end{document}